\documentclass[]{emulateapj}
\usepackage{natbib}
\usepackage{multirow}
\def\ha{H$\alpha$}
\def\hb{H$\beta$}
\def\sfre{$SFR_{e}$}
\def\sfrha{$SFR_{{\rm H}\alpha}$}
\def\mi{$\mu$m}
\def\myd4n{D$_n$(4000)}
\def\ltir{$L_{TIR}$}
\def\lhaobs{$L_{{\rm H}\alpha}^{obs}$}
\def\lhacorr{$L_{{\rm H}\alpha}^{corr}$}
\def\lne{$L_{\rm Ne}$}
\newcommand\newion[2]{#1$\;${\scshape{#2}}}
\def\hii{\newion{H}{ii}}
\def\neii{[\newion{Ne}{ii}]12.8$\mu$m}
\def\neiii{[\newion{Ne}{iii}]15.5$\mu$m}
\def\siii{[\newion{S}{iii}]18.7$\mu$m}
\def\oiv{[\newion{O}{iv}]25.9$\mu$m}
\def\si2{[\newion{Si}{ii}]34.8$\mu$m}

\def\myo3hb{[\newion{O}{iii}]$\lambda 5007$/H$\beta$}
\def\n2ha{[\newion{N}{ii}]$\lambda 6583$/H$\alpha$}

\shorttitle{MIR indicators of SF and AGN in Normal Galaxies}
\shortauthors{Treyer et al.}

\begin{document}

\title{Mid-Infrared Spectral Indicators of Star-Formation and AGN Activity in Normal Galaxies}

\author{Marie Treyer\altaffilmark{1}, David Schiminovich\altaffilmark{2}, Benjamin D. Johnson\altaffilmark{3}, Matt O'Dowd\altaffilmark{2},
Christopher D. Martin\altaffilmark{1}, Ted Wyder\altaffilmark{1}, St\'ephane Charlot\altaffilmark{4}, Timothy Heckman\altaffilmark{5}, 
Lucimara Martins\altaffilmark{6}, Mark Seibert\altaffilmark{7}, J. M. van der Hulst\altaffilmark{8}}

\altaffiltext{1}{California Institute of Technology, MC 278-17, 1200 E. California Boulevard, Pasadena, CA 91125, USA; treyer@srl.caltech.edu}
\altaffiltext{2}{Astronomy Department, Columbia University, 550 W. 120 St., New York, NY 10027,USA}
\altaffiltext{3}{Institute of Astronomy, University of Cambridge, Madingley Road, Cambridge CB3 0HA, UK}
\altaffiltext{4}{Institut d'Astrophysique de Paris, UMR 7095, 98bis Bvd Arago, 75014 Paris, France}
\altaffiltext{5}{Department of Physics and Astronomy, Johns Hopkins University, Homewood Campus, Baltimore, MD 21218, USA}
\altaffiltext{6}{NAT - Universidade Cruzeiro do Sul, Rua Galv\~ ao Bueno, 868, S\~ ao Paulo, SP, 01506-000, Brazil}
\altaffiltext{7}{Observatories of the Carnegie Institution of Washington, 813 Santa Barbara Street, Pasadena, CA 91101, USA}
\altaffiltext{8}{Kapteyn Astronomical Institute, University of Groningen, Netherlands} 

\begin{abstract}
We investigate the use of mid-infrared (MIR) polycyclic aromatic hydrocarbon (PAH) bands, 
continuum and emission lines as probes of star-formation and active galactic nuclei (AGN) activity 
in a sample of 100 `normal' and local ($z \sim 0.1$) emission-line galaxies. 
The MIR spectra were obtained with the {\it Spitzer} Space Telescope Infrared Spectrograph (IRS) 
as part of the {\it Spitzer}-SDSS-GALEX Spectroscopic Survey (SSGSS) which includes multi-wavelength
photometry from the ultraviolet to the far-infrared and optical spectroscopy. The continuum and
features were extracted using PAHFIT (Smith et al. 2007), a decomposition code which we find to 
yield PAH equivalent widths up to $\sim 30$ times larger than the commonly used spline methods. 
Despite the lack of extreme objects in our sample (such as strong AGNs, low metallicity galaxies 
or ULIRGs), we find significant variations in PAH, continuum and emission line properties 
and systematic trends between these MIR properties and optically derived physical 
properties such as age, metallicity and radiation field hardness.
We revisit the diagnostic diagram relating PAH equivalent widths and \neii/\oiv\ line ratios and find it to be 
in much better agreement with the standard optical star-formation/AGN classification than when spline decompositions are used, 
while also potentially revealing obscured AGNs. 
The luminosity of individual PAH components, of the continuum,
and with poorer statistics, of the neon emission lines and molecular hydrogen lines,
are found to be tightly correlated to the total infrared luminosity,
making individual MIR components good gauges of the total dust emission in SF galaxies.
Like the total infrared luminosity, these individual components can be used to estimate 
dust attenuation in the UV and in \ha\ lines based on energy balance arguments.
We also propose average scaling relations between these components and dust corrected, \ha\ derived 
star-formation rates. 
\end{abstract}

\section{Introduction}
Determining the main source of ionizing radiation and the star-formation rate (SFR) of galaxies are essential quests in the study of galaxy evolution. 
While optical diagnostic diagrams \citep[e.g.][]{BPT1981} allows a rather clear distinction between star-formation (SF) and accretion disk processes, 
they are limited  to - by definition - visible components and are at this point extremely difficult to apply at high redshifts. 
The same caveats apply to the 
measurement of SFRs from optical lines. Mid-infrared (MIR) spectroscopy offers a potent alternative, much less sensitive to 
interstellar exinction. MIR galaxy spectra exhibit an array of features 
arising essentially from 
(1) a continuous distribution of dust grains, the smallest of which (VSGs for Very Small Grains) produce the continuum longward 
of $\sim 10\mu m$ \citep{Desert_etal1990} while larger ones containing silicates produce absorption features at 9.7 and 18\mi\ \citep{LebofskyRieke1979};
(2) ionized interstellar gas producing fine-structure lines; 
and (3) molecular gas producing most notably a series of broad emission features, most prominent in the $6-17$\mi\ range, which were
previously referred to as ``Unidentified Infrared Bands'' but are now commonly attributed to vibrational emission of large polycyclic 
aromatic hydocarbon (PAH) molecules \citep{LegerPuget1984, Allamandola_etal1985, PugetLeger1989}. Rotational lines of molecular hydrogen
are also detected \citep[and references therein]{Roussel_etal2007}.
MIR diagnostics have been devised to unveil the ionizing source heating these components
\citep[e.g.][]{Voit1992_diagnostics,Genzel_etal1998, Laurent_etal2000, Spoon_etal2007}
and calibrations have been proposed to derive SFRs from their luminosities \citep[e.g.][]{HoKeto2007, Zhu_etal2008, Rieke_etal2009, HernanCaballero_etal2009}. 
As these calibrations and the resolving power of the various diagnostic diagrams vary with galaxy types, it is important to review the MIR spectral properties of 
well defined classes of objects. The Infrared Spectrograph (IRS) on board the {\it Spitzer} satellite has allowed many such investigations, building on earlier fundamental 
results from the Infrared Space Observatory \citep{CesarskySauvage1999, GenzelCesarsky2000}. Much attention has been devoted to extreme sources such as ULIRGs 
\citep{Armus_etal2007, Farrah_etal2007, Desai_etal2007}, 
starburst galaxies \citep{Brandl_etal2006}, 
AGNs \citep{Weedman_etal2005, Deo_etal2009, Thompson_etal2009} or 
QSOs \citep{Cao_etal2008}. 
IRS observations of the SINGS sample \citep{SINGS2003} have also provided many new results about the central region of nearby galaxies 
spanning a broad range of physical properties \citep{Dale_etal2006, Smith_etal2007, Dale_etal2009}. 
However few studies have yet focused on `normal' galaxies. Still, questions remain open on this seemingly unexciting class of objects.  

Whether VSG or PAH emission can be used to trace SF in normal galaxies has been often debated in recent years 
\citep{Roussel_etal2001, ForsterSchreiber_etal2004, Peeters_etal2004, Calzetti_etal2007, Kennicutt_etal2009}. 
Resolved observations of star-forming regions have shown that the VSG continuum strongly peaks inside \hii\ regions
while PAH features dominate in photodissociation regions (PDRs) and get weaker nearer the core of \hii\ regions, where the
molecules are thought to be destroyed by the intense radiation fields 
\citep[e.g.][]{Boulanger_etal1988, Giard_etal1994, Cesarsky_etal1996, Verstraete_etal1996, Povich_etal2007,Gordon_etal2008}.
However neutral PAH emission has recently been reported inside \hii\ region \citep{Compiegne_etal2007}.
They are also found in the interstellar medium (ISM), indicating that they must also be excited by softer 
near-UV or optical photons \citep[e.g.][]{LiDraine2002, Calzetti_etal2007}, making them perhaps better tracer of B stars than of SF \citep{Peeters_etal2004}.
VSG emission is also observed in the ISM but with higher PAH/VSG surface brightness ratios than in SF regions \citep{Bendo_etal2008}.
Despite much complexity on small scales however, integrated MIR luminosities at 24\mi\ and 8\mi\, tracing the VSG and PAH emissions
respectively, are found to correlate with \ha\ luminosities \citep[e.g.][]{Zhu_etal2008}, though not linearly and with scatter 
\citep{Kennicutt_etal2009} leading to uncertain SFR estimates. 
 
An additional source of uncertainty is the common occurence of AGN in normal galaxies. 
PAH molecules are also  thought to get destroyed near the hard radiation fields of AGNs \citep{DesertDennefeld1988, Voit1992_pahdestruction} 
however not totally and as was shown recently from IRS spectroscopy, preferentially at short wavelengths \citep{Smith_etal2007, ODowd_etal2009}. 
There is in fact no a priori reason why PAH emission could not be excited by UV photons from an AGN \citep{Farrah_etal2007}.
This further compromises the use of PAH bands as SFR indicators, unless AGNs can be reliably detected in the MIR spectra of normal galaxies. 

We have obtained IRS spectra for a sample of 101 normal galaxies at $z\sim 0.1$ with the goal to tackle the above issues, making use
of additional multi-wavelength (ultraviolet to far-infrared) photometric data and optical spectroscopic data.
The first results of this survey have been reported by \cite{ODowd_etal2009} who analyzed the dependence of the
relative strength of PAH emission features with optical measures of SF and AGN activity. 
We are pursuing this study by comparing optical and MIR diagnostic diagrams to detect AGN presence in normal galaxies and by investigating 
the use of PAH, MIR continuum and emission line luminosities as tracer of the total IR luminosity and \ha\ derived SFRs. 
The sample, IRS data and spectral decomposition method are described in section 2. Section 3 presents the continuum, PAH and emission line properties
of the galaxies as a function of SF and AGN activity. In particular we analyze the dependency of PAH equivalent widths with age, metallicity and 
radiation field hardness, as well as the efficiency of MIR diagnostics to detect optically classified AGNs in these galaxies.
We present correlations between the luminosities of MIR components and the total IR luminosity in section 4 and between these components
and SFR estimates in section 5. Our conclusion are summarized in section 6.
Throughout the paper we assume a flat $\Lambda$CDM cosmology with $H_0=70~{\rm km~ s^{-1} Mpc^{-1}}$, 
$\Omega_M=0.3$ and $\Omega_{\Lambda}=0.7$, and a Kroupa IMF \citep{KroupaIMF} for SFR calibrations.

\begin{figure*}[tb]
\plottwo{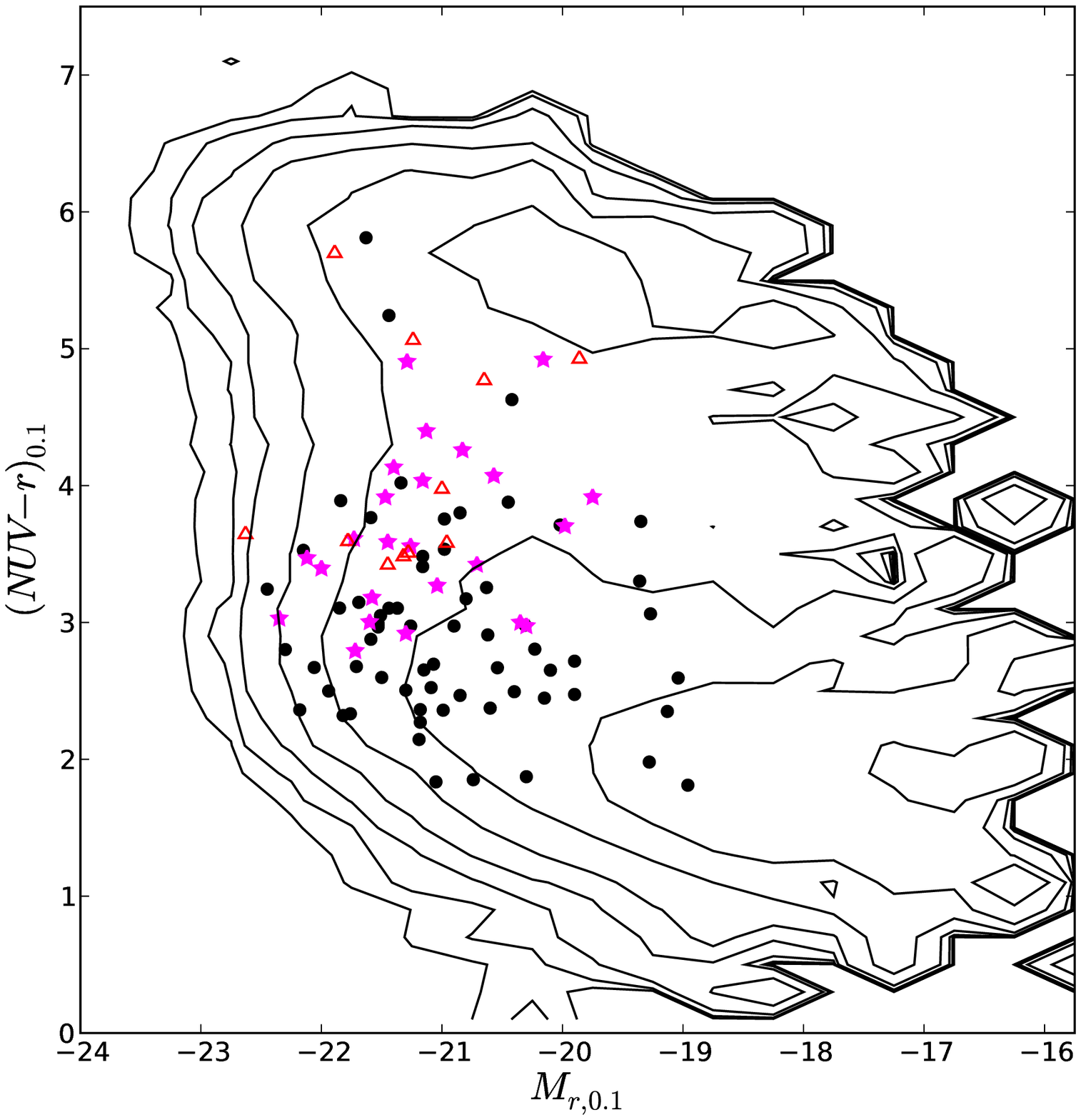}{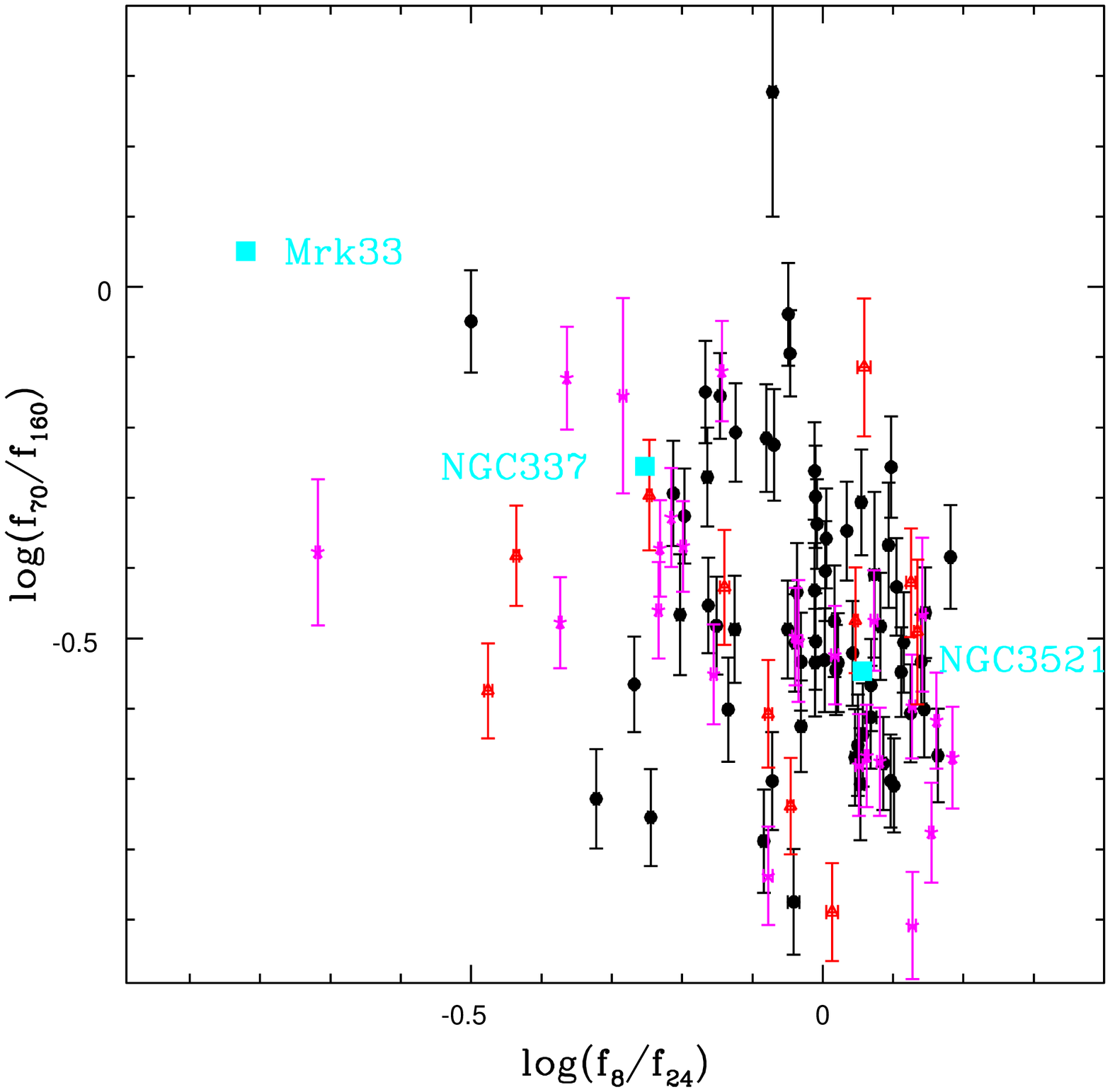}
\caption{{\it Left:} $NUV-r$ color vs $r$-band magnitude diagram showing the location of SSGSS galaxies with respect to the underlying 
local population shown as volume density contours \citep{Wyder_etal2007}. Galaxies separate into two well-defined blue and red sequences. 
In this and all following figures, star-forming galaxies are represented as black dots, composite galaxies as pink stars and AGNs 
as open red triangles. The SSGSS sample is dominated by blue sequence galaxies with a small fraction (mostly AGNs) on the red sequence.
{\it Right:} Infrared color-color diagram: $f_8/f_{24}$ versus $f_{70}/f_{160}$ flux ratios. `Normal' quiescent SF galaxies are found 
towards the bottom right corner (stronger PAH emission and cooler dust in the FIR) while starburst galaxies populate the top left corner 
(strong hot dust continuum in the MIR and warm dust emission in the FIR) .
The {\it Spitzer} data for NGC 35321, NGC 337 and Mrk 33 are taken from \cite{Dale_etal2007}. 
\label{fig:cmd}
}
\end{figure*}

\section{The SSGSS Sample}

The {\it Spitzer}-SDSS-GALEX Spectroscopic Survey (SSGSS) is a MIR spectroscopic survey of 101 local star-forming
galaxies using the Infrared Spectrograph (IRS) \citep{Houck_etal2004} aboard the {\it Spitzer} Satellite. The IRS
and corollary data are available at: http://www.astro.columbia.edu/ssgss/.

\subsection{The parent sample}

The sample is drawn from the Lockman Hole region which has been extensively surveyed at multiple wavelengths. 
In particular UV photometry from GALEX (1500 and 2300\AA), optical imaging and spectroscopic observations from SDSS and 
infrared photometry (IRAC and MIPS channels) from {\it Spitzer} (SWIRE) are available for all SSGSS galaxies. 
The redshifts span $0.03<z<0.21$ with a mean of 0.09 similar to that of the full SDSS spectroscopic sample. 
The sample has a surface brightness limit of 0.75 MJy sr$^{-1}$ at 5.8\mi\ and a flux limit of 1.5mJy at 24\mi. Due to 
these cuts the sample does not contain very low mass/low metallicity/low extinction galaxies ($9.3\le M_\odot \le 11.3$, 
$8.7\le {\rm log(O/H)}+12\le 9.2$ and $0.4<A_{{\rm H}\alpha}<2.3$) but it was selected
to cover the range of physical properties of `normal' galaxies. 

These galaxies are divided up into 3 categories: star-formation dominated galaxies (referred to as `SF galaxies'), composite galaxies 
(SF galaxies with an AGN component) and AGN dominated galaxies, according to their location on the ``BPT diagram'' \citep{BPT1981}, 
which shows \n2ha\ (a proxy for gas phase metallicity) against  \myo3hb\ 
(a measure of the hardness of the radiation field).
AGNs (all Seyfert 2's in our sample) are isolated by the theoretical boundary of \cite{Kewley_etal2001}
while SF galaxies and composite galaxies are separated by the empirical boundary of \cite{Kauffmann_etal2003}.
In all following figures, SF galaxies are represented as black dots, composite galaxies 
as pink stars and AGNs as open red triangles. We note that this optical classification may miss obscured AGNs
contributing to the MIR emission. 

The location of the sample in the
$NUV-r$ color versus $r$-band absolute magnitude diagram is shown in Figure \ref{fig:cmd} (left panel) with the volume
density contours of the underlying local population \citep{Wyder_etal2007}. 
Galaxies in this diagram separate into two well-defined blue and red sequences that become redder with increasing luminosity. 
The red sequence tend to be dominated by high surface brightness, early type galaxies with low ratios of current to past averaged 
SF, while the blue sequence is populated by morphologically late-type galaxies with lower surface brightness and on-going SF activity
\citep[e.g.][]{Strateva01}. The color variation along the blue sequence is due to a combination of dust, SF history 
and metallicity \citep{Wyder_etal2007}. Unsurprisingly given its selection criteria, our sample is dominated by blue sequence galaxies, 
although a small fraction (mostly AGNs) are found on the red sequence.

The right panel of Figure \ref{fig:cmd} shows the distribution of the sample in the  $f_{8}$/$f_{24} - f_{70}$/$f_{160}$
plane, where $f_{8}$/$f_{24}$ is the 8\mi\ to 24\mi\ restframe flux ratio (4$^{th}$ IRAC band to 1$^{rst}$ MIPS band) and 
$f_{70}$/$f_{160}$ is the 70\mi\ to 160\mi\ restframe flux ratio (2$^{nd}$ to 3$^{rd}$ MIPS bands). The infrared $k$-corrections are described
in Section 4. This figure illustrates the range of IR properties covered by the sample (\cite{DaCunha_etal2008} and references therein). 
Starburst galaxies tend to populate the top left corner (strong hot dust continuum in the MIR and warm dust emission in the FIR) 
while more quiescent SF galaxies move down towards the bottom right corner (stronger PAH emission and cooler 
dust in the FIR). The {\it Spitzer} data for NGC 35321, NGC 337 and Mrk 33 are from \cite{Dale_etal2007}.

\begin{figure}
\plotone{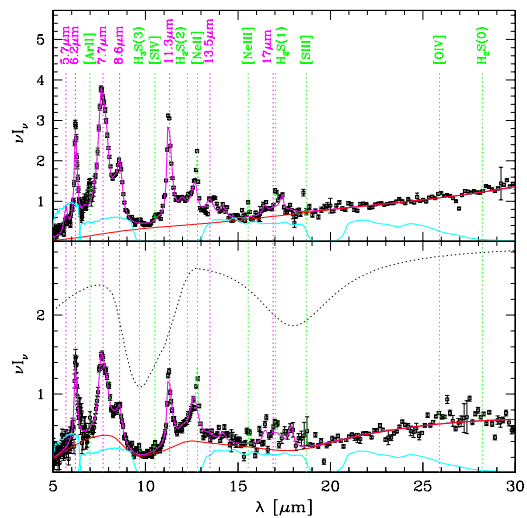}
\caption{Two example spectra with best fits ($\nu I_{\nu}$ on the $y-$axis in units of  $10^{11} {\rm Jy~Hz}$). 
The fits are outputs of PAHFIT \citep{Smith_etal2007}.
The red lines fit the continuum, the purple lines fit the PAH features and the dotted green lines fit the 
emission lines. The main PAH features are indicated in purple and the main emissions lines in green.
The blue curves show - from left to right --- the filter responses of the IRAC bands at 6 and 8\mi, 
of the IRS blue Peak Up band at 16\mi\ and of the MIPS band at 24\mi\ bands.  
The galaxy in the top panel is a SF galaxy with no silicate absorption ($\tau_{\lambda}=0$); 
the galaxy in the bottom panel is an AGN with strong silicate absorption features at 9.7\mi\ and 18\mi\  ($\tau_{\lambda}=2.5$).
The extinction $(1-e^{-\tau_{\lambda}})/\tau_{\lambda}$ is shown as the dotted line in arbitrary units. 
\label{fig:exspectra}
}
\end{figure}

\subsection{The IRS Spectra}

Low resolution spectroscopic observations were acquired for the full sample using the Short-Low (SL) and 
Long Low (LL) IRS modules, covering 5\mi\ to 38\mi\ with resolving power $\sim$ 60 to 127. High resolution 
spectra were obtained for the 33 brightest galaxies using the Short-High (SH) IRS module, covering 
10\mi\ to 19.6\mi\ with a resolving power of $\sim 600$. A detailed description of the data acquisition
and reduction can be found in O'Dowd et al. (in preparation). In short, standard IRS calibrations were performed 
by the {\it Spitzer} Pipeline version S15.3.0 (ramp fitting, dark substraction, droop, linearity correction,
distortion correction, flat fielding, masking and interpolation, and wavelength calibration). Sky subtraction
was performed manually with sky frames constructed from the 2-D data frames, utilizing the shift in galaxy
spectrum position between orders to obtain clean sky regions. IRSCLEAN (v1.9) was used to clean bad and rogue
pixels. SPICE was used to extract 1-D spectra, which were combined and stitched manually by weighted mean.  
After rejecting problematic data, the final sample consists of 82 galaxies (56 SF galaxies, 19 composite galaxies
and 7 AGNs) with low resolution spectra, of which 31 (23 SF galaxies, 6 composite galaxies and 2 AGNs) 
have high resolution spectra as well. 

\subsection{PAHFIT decomposition}

We use the PAHFIT spectral decomposition code (v1.2) of \cite{Smith_etal2007} (hereafter S07) to fit each spectrum
as a sum of dust attenuated starlight continuum, thermal dust continuum, PAH features and emission lines. The 
absorbing dust is assumed to be uniformly mixed with the emitting material. The code performs a $\chi^2$
fitting of the emergent flux as the sum of the following components (Eq. 1 in S07): 
\begin{equation}
I_\nu=\left[ \tau_{\star}B_{\nu}(T_{\star})+\sum_{m=1}^M \tau_m {B_{\nu}(T_m) \over (\lambda/\lambda_0)^2}
+\sum_{r=1}^R I_r(\nu)\right]{(1-e^{-\tau_{\lambda}})\over \tau_{\lambda}},
\label{eq:pahfit}
\end{equation}

where $B_{\nu}$ is the backbody function, $T_{\star}=5000~K$ is the temperature of the stellar continuum,
$T_m={35,40,50,65,90,135,200,300~K}$ are 8 thermal dust continuum temperatures, the $I_r(\nu)$ consist
of 25 PAH emission features modeled as Drude profiles and 18 unresolved emission lines modeled as Gaussian profiles, 
and $\tau_{\lambda}$ is the dust opacity, normalized at $\lambda_0=9.7\mu m$. The specifics of these components are 
described in S07.  The Drude profile, which has more power in the
extended wings than a Gaussian, is the theoretical profile for a classical damped harmonic oscillator 
and is thus a natural choice to model PAH emission. Some of the PAH features are modeled by several 
blended subfeatures, most prominently the PAH complex at 7.7\mi\ which is modeled by a combination of
3 Drude profiles and the PAH complex at 17\mi\ modeled by 4 such profiles.  
The continuum components have little significance individually, it is their combination 
that is meant to produce a physically realistic continuum. We find that the stellar continuum 
is negligible for most galaxies, which is probably not surprising since it is practically unconstrained.  
The theoretical value of $0.232 \times f_{3.6\mu m}$ \citep{Helou_etal2004} for the stellar contribution to the 8\mi\ band
is $\sim 20\%$, however it is probably an upper limit since the 3.6\mi\ flux may be contaminated by the 3.3\mi\ PAH feature 
for galaxies at $z\sim 0.1$. We also find that silicate absorption is negligible ($\tau_{9.7}<0.1$) for 63\% of the sample, 
however 3 out of 7 AGNs are among the galaxies showing the strongest absorption features.
From the best fit decompositions\footnote{corrected for the PAHFIT v1.2 $(1+z)$ overestimate.} 
we compute the fluxes of the PAH features, emission lines and continuum at various points corrected for silicate absorption 
as well as the total restframe and observed fluxes in the MIR {\it Spitzer} bands. 
The fluxes of the main MIR components used in this paper are listed in Table \ref{table:fluxes}.

Figure \ref{fig:exspectra} shows two examples of our IRS spectra with best fit decomposition from PAHFIT (restframe).
The galaxy in the top panel is a typical SF galaxy (\#30); that in the bottom panel is an AGN (\#93) 
showing the strongest silicate absorption features at 9.7\mi\ and 18\mi\ in the sample ($\tau_{9.7}=2.5$). 
The extinction $(1-e^{-\tau_{\lambda}})/\tau_{\lambda}$ is shown as the dotted line in arbitrary units.

\begin{figure}
\plotone{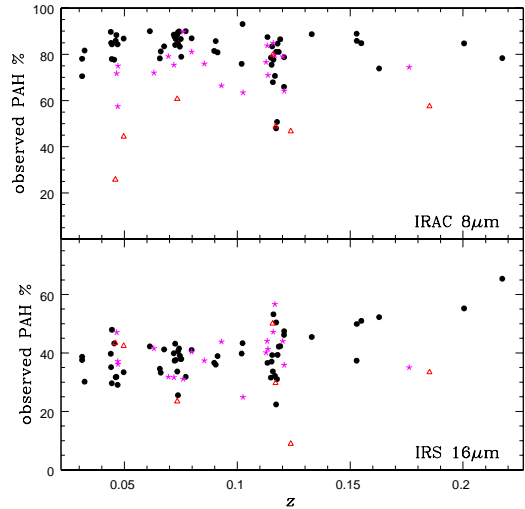} 
\caption{Observed PAH fractions in the 8\mi\ IRAC band and 16\mi\ IRS band as a function 
of redshift. Symbols are as described in Figure \ref{fig:cmd}. 
The 8\mi\ flux is dominated by PAH emission for most galaxies while the 16\mi\ flux is dominated 
by continuum emission except for the two highest redshift sources. 
\label{fig:pahfraction} 
}
\end{figure} 

Figure \ref{fig:pahfraction} shows the observed fractions of PAH emission in the 8\mi\ IRAC band (often used as
a proxy for PAH emission) and in the 16\mi\ IRS band as a function of redshift.
For our local sample the 8\mi\ IRAC channel picks up the 7.7\mi\ PAH complex, plus the 6.2\mi\ PAH feature for galaxies 
at $z>0.05$ and the 8.6\mi\ PAH feature for galaxies at $z<0.05$. Both the observed and restframe fluxes in this band
are largely dominated by PAH emission for most SF and composite galaxies. The continuum dominates only for 1 AGN.
The 16\mi\ IRS Peak-Up band collects photons from the 17\mi\ PAH complex  
plus other smaller PAH features around 14\mi\ and the large 12.7\mi\ complex for galaxies at $z>0.06$. The observed 
16\mi\ flux includes more PAH emission than the restframe flux, which is continuum dominated ($\sim 70\%$) for all galaxies.
The observed 24\mi\ MIPS channel is vastly dominated by the continuum for all sources. The highest PAH contribution (18\%)
comes from the redshifted 18.92\mi\ PAH feature and the red wing of the 17\mi\ PAH feature for the highest redshift object 
($z=0.217$). PAH emission starts to dominate the 24\mi\ channel for galaxies at $z>1$.

\begin{figure}
\plotone{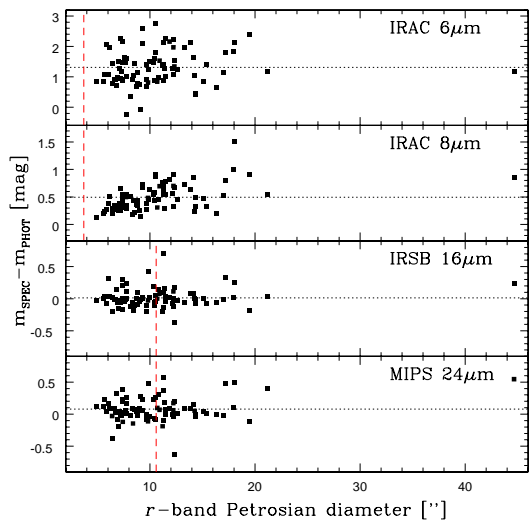}
\caption{Aperture corrections in magnitude at 6, 8, 16 and 24\mi\ as a function of $r$-band Petrosian diameter in arcseconds. 
The dotted lines mark the average aperture corrections. The vertical lines show the slit widths of the SL and LL 
modules (upper and lower panels respectively). 
\label{fig:apcorr-rad}
}
\end{figure}

\subsection{Aperture corrections}

The SL and LL IRS modules have slit widths of 3.6'' and 10.5'' respectively, while the mean angular size of the
sample is 10''.
The corrections applied to stitch the two modules together in the overlap region ($14.0-14.5\mu m$) are explained 
in detail by O'Dowd et al. (in preparation).
Wavelength dependent aperture effects also arise from the wavelength dependent PSF 
(increased sampling of the central regions of extended galaxies with increasing wavelength).
To remedy these effects, we compute spectral magnitudes in the MIR {\it Spitzer} bands
both from the data and from the PAHFIT SEDs using the {\it Spitzer Synthetic Photometry} cookbook. 
We find excellent agreement between data and fits except in the 6\mi\ IRAC bands, the data being noisy and
the fits unreliable below 5.8\mi. For this reason we do not make use of fluxes in this part of the spectrum. 
The difference between the PAHFIT spectral magnitudes and the photometric magnitudes are used as 
aperture corrections at the effective wavelengths. These corrections are shown in Figure \ref{fig:apcorr-rad}
as a function of $r$-band petrosian diameter and listed in Table \ref{table:aper}. The vertical lines in the upper
and lower panels show the slit widths of the SL and LL modules respectively. It is clear that flux is lost at 8\mi,
however at longer wavelengths we do not find that much flux is lost even when the optical petrosian diameter is larger 
than the slit width, which we attribute to the larger PSF.
Corrections at intermediate wavelengths are obtained by interpolation. The mean corrections are $\sim 1.2$ mag at 6\mi, 
$\sim 0.5$ mag at 8\mi\ and $< 0.1$ at 16 and 24\mi.
In the following, all MIR luminosities computed from the PAHFIT decomposition (PAH, continuum and emission line luminosities 
as well as total restframe luminosities in the {\it Spitzer} bands) are corrected for aperture as described in this section. 

\begin{figure}
\plotone{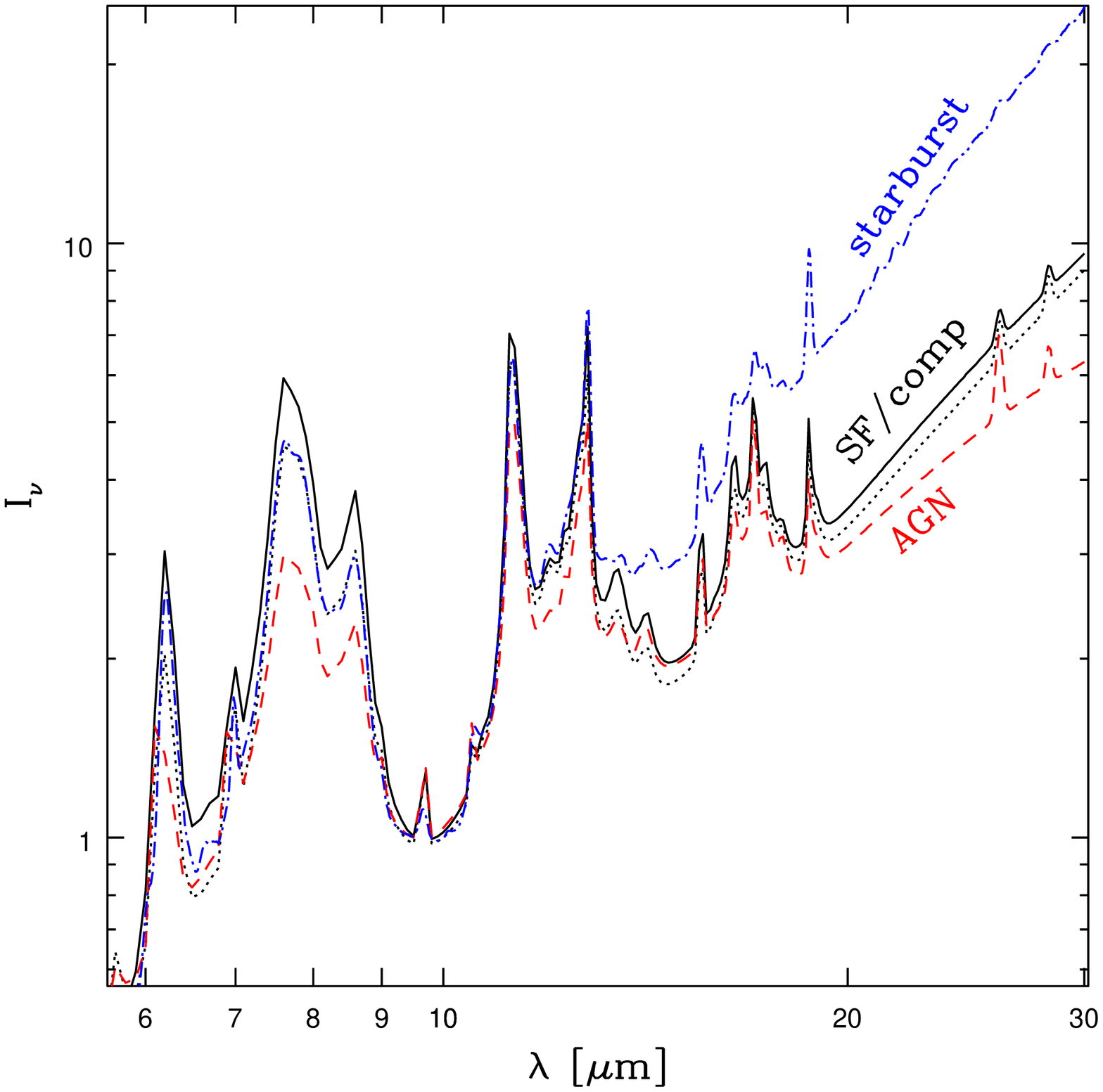}
\caption{The mean spectra of SF galaxies (solid line), composite galaxies (dotted line) and AGNs (dashed line) 
normalized at 10\mi. The dot-dashed spectrum is the average starburst spectrum of \cite{Brandl_etal2006}.
The transition from starburst to SF galaxy to AGN is marked by a declining continuum slope, decreased \neii\ and \siii\  
and enhanced \oiv. The AGN and starburst spectra also show depleted PAH emission at low wavelength compared to 
the SF spectrum.
\label{fig:meanspec}
}
\end{figure}

\section{MIR Spectral properties}

Figure \ref{fig:meanspec} shows the mean spectra of our SF galaxies (solid line), composite galaxies 
(dotted line) and AGNs (dashed line) as well as the average starburst spectrum of \cite{Brandl_etal2006} (dot-dash), 
normalized at 10\mi.
The transition from starburst to SF galaxy to AGN is associated with a declining continuum slope,
most dramatic between the starburst spectrum and the normal SF spectrum. Indeed \hii\ regions and starburst galaxies are found to 
exhibit a steep rising VSG continuum component longward of $\sim 9\mu m$ \citep[e.g.][]{Cesarsky_etal1996, Laurent_etal2000, 
Dale_etal2001, Peeters_etal2004}. The transition is also marked by
decreased \neii\ and \siii\ line emission and enhanced \oiv\ line emission \citep[e.g.][]{Genzel_etal1998}.
The AGN spectrum, and to a lesser extent the starburst spectrum, show weaker PAH emission at low wavelength than 
the SF spectrum, an effect attributed to the destruction of PAHs in intense far-UV radiation fields. 

\subsection{PAH features, continuum and emission lines}

\begin{figure}
\plotone{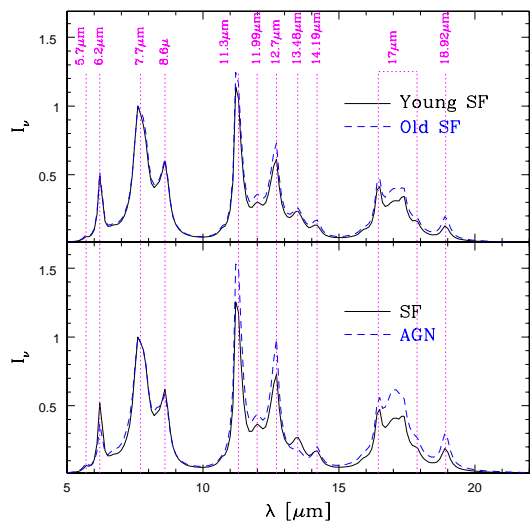}
\caption{{\it Top:} The mean PAH component of `young' SF galaxies with $1.1<D_n(4000)<1.3$  (solid line) and `old' SF galaxies 
with $1.3<D_n(4000)<1.6$ (dashed line). 
{\it Bottom:} The mean PAH component of SF galaxies (solid line) and AGNs (dashed line) in the $1.3<D_n(4000)<1.6$ range 
where both types have similar mean \myd4n $\sim 1.4$.
The spectral components are normalized by the peak intensity of the 7.7\mi\ feature.   
The main difference between the two pairs is enhanced PAH emission at large wavelengths with respect to the 7.7\mi\ feature.
\label{fig:specdiff-pah}
}
\end{figure}

PAHFIT allows us to compare the different spectral components of different galaxy types separately. 
The top panel of Figure \ref{fig:specdiff-pah} shows the average PAH component of `young' SF galaxies 
with  $1.1<D_n(4000)<1.3$ ($<D_n(4000)>=1.2$) and that of `old' SF galaxies with $1.3<D_n(4000)<1.6$ ($<D_n(4000)>=1.4$).
The 4000\AA\ break \myd4n \citep{Balogh98} is a measure of the average age of the stellar populations.
The separating value is simply the median of the distribution. 
The bottom panel shows the average PAH components of SF galaxies and AGNs in the $1.3<D_n(4000)<1.6$ range where both 
types have similar mean \myd4n\ $\sim 1.4$ (there are only 4 AGNs in that bin). 
All spectra are normalized by the peak intensity of the 7.7\mi\ feature. 
The main difference between the pairs in both panels is enhanced PAH emission at large wavelengths with respect to the 7.7\mi\ feature,
i.e. an increase in the ratio of high to low wavelength PAHs associated with both AGN presence and increased stellar population age. 
This increase is most pronounced in the lower panel (AGN versus SF) where a decrease in the 6.2\mi\ feature with respect to the 7.7\mi\ feature
is also noticeable. 
The variations in PAH ratios in this sample have been thoroughly studied by \cite{ODowd_etal2009} and shown to be
statistically significant. 
 These variations are much more dramatic for AGNs with harder radiation fields than
those in the present sample (e.g. S07, their Figure 14). They can be attributed to a change in the fraction of neutral 
to ionized PAHs responsible for the high and low wavelength features respectively, and/or to the destruction by 
hard radiation fields in AGNs of the smallest PAH grains emitting at low wavelengths  (S07, and references therein). 
The variations of PAH strengths with age, metallicity and radiation field hardness are explored in more detail in the next section.

\begin{figure}
\vspace{-2.cm}
\plotone{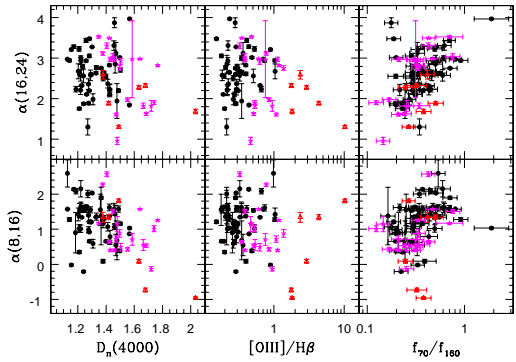}
\caption{MIR color indices $\alpha(8,16)$ and  $\alpha(16,24)$  as a function of \myd4n, \myo3hb\ ratio and $f_{70}/f_{160}$ 
restframe colors. Symbols are as described in Figure \ref{fig:cmd}. 
Although older galaxies and AGNs have shallower continuum slopes on average, 
little correlation is found with \myd4n\ nor radiation field hardness. 
The correlation with FIR color for SF galaxies may reflect a sequence in the peak wavelength
of the dust SED: the MIR slope steepens at it gets closer to the peak while $f_{70}/f_{160}$ increases.
\label{fig:slope-x}}
\end{figure}

We define the continuum slope or MIR color index between wavelengths $\lambda_1$ and $\lambda_2$ as:
\begin{equation}
\alpha(\lambda_1,\lambda_2)={ {\rm log}\left[I_\nu^{cont}(\lambda_2)/I_\nu^{cont}(\lambda_1) \right] \over {\rm log}(\lambda_2/\lambda_1) }
\label{eq:slope}
\end{equation}
where $I_\nu^{cont}(\lambda)$ is the continuum component of Eq. \ref{eq:pahfit} at $\lambda$ corrected for silicate absorption. 
This would be the index $\beta$ of a continuum spectrum of the form $I_\nu\propto \lambda ^{\beta}$. 
Figure \ref{fig:slope-x} shows $\alpha(8,16)$ and $\alpha(16,24)$ as a function of \myd4n, \myo3hb\ 
and the restframe  $f_{70}/f_{160}$ color (see Section 4 for details on the $k$-corrections).
As discussed above and shown in Figure \ref{fig:meanspec}, 
the mean MIR slope is found to steepen from quiescent galaxies to starburts of increasing activity \citep{Dale_etal2001}
and to be shallower for AGNs \citep[e.g.][]{GenzelCesarsky2000}. However our
indices span a significant range ($\sim$3 dex) with little correlation with the age of the stellar populations or radiation field hardness.
Older galaxies (\myd4n $>1.6$) do tend to populate the low end of the distribution  (i.e. have shallower slopes) in both cases, 
as do AGNs in the red part of the spectrum however a flatter continuum could not be used as a criterion to separate 
AGNs from SF galaxies, as previously reported by \cite{Weedman_etal2005}. 
The correlation with FIR color for SF galaxies is more striking, especially at longer MIR wavelengths. This may be expected if the peak 
of the dust SED (a blackbody modified by the emissivity) is located shortward of $\sim$ 100\mi. In this case as the peak wavelength 
decreases, the MIR continuum slope gets closer to the peak and therefore steepens while  $f_{70}/f_{160}$ increases.  

\begin{figure}
\plotone{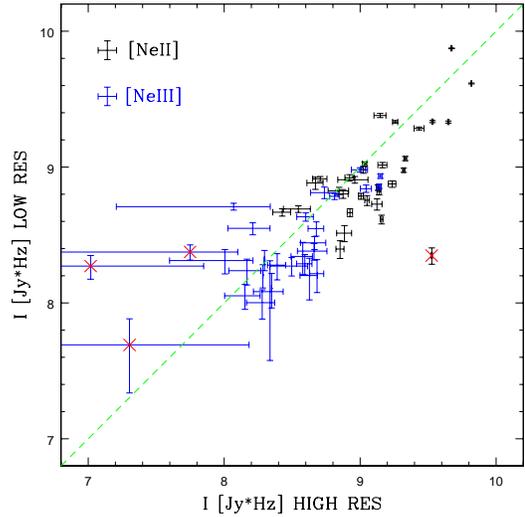}
\caption{Comparison between the low and high resolution line fluxes of \neii\ (in black) and \neiii\ (in blue)
for a subsample of 31 galaxies. Excluding the objects marked with a red cross, the rms of the correlation is 0.22. 
\label{fig:lr-hr}
}
\end{figure}

Finally we look at variations in the emission line components.
The lines modeled by PAHFIT in the low resolution spectra are meant to provide a realistic decomposition of 
the blended PAH features and the continuum (S07) but the spectral resolution is of the same order as the FWHM of the lines. 
Figure \ref{fig:lr-hr} shows the comparison between the high and low resolution fluxes of the \neii\ and \neiii\ lines (black and blue error
bars respectively) for the subsample observed with the SH module. The high resolution lines were also measured using PAHFIT with the
default settings. We make no attempt at aperture correction on this plot. 
Excluding the 3 extreme error bars among the \neiii\ fluxes at high resolution and the outlier among the \neii\ fluxes
(marked as red crosses in Figure \ref{fig:lr-hr}), the fitting procedure at low resolution recovers 
the high resolution fluxes with an rms of 0.22 dex, a reasonable estimate considering the factor of 10 difference in spectral resolution.
In particular the PAH contamination for the \neii\ line does not seem to be a significant problem in the SL data using PAHFIT.
For the purpose of the present statistical analysis we use the low resolution line 
measurements which are available for the full sample and over the full range of wavelengths.
We refer to O'Dowd et al. (in preparation) for a detailed comparison between the high and low resolution data.
 
\begin{figure}
\plotone{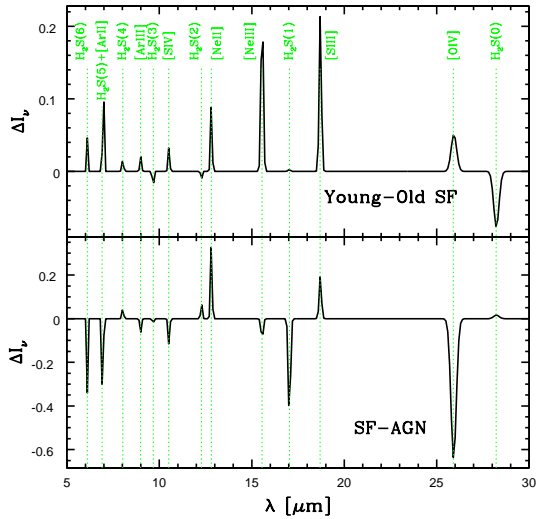}
\caption{{\it Top:} The difference - $\Delta I_{\nu}$ - between the mean emission line component of `young' SF galaxies 
with $1.1<D_n(4000)<1.3$ and that of `old' SF galaxies with $1.3<D_n(4000)<1.6$. 
{\it Bottom:} The difference between the mean emission line component of SF galaxies and that of AGNs in the range $1.3<D_n(4000)<1.6$. 
The spectral components are normalized by the total flux in the 16\mi\ IRS band.
The most significant features are decreased \neiii\ and \siii\ and increased H$_2S(0)$ 
in older SF galaxies with respect to younger ones, and increased \oiv\ and H$_2S(1)$ in AGNs along 
with diminished \neii\ and \siii\ with respect to SF galaxies. 
\label{fig:specdiff-el}
}
\end{figure}

The top panel of Figure \ref{fig:specdiff-el} shows the difference -- $\Delta I_{\nu}$ -- between the average emission line component of `young' 
SF galaxies ($<D_n(4000)>=1.2$) and that of `old' SF galaxies ($<D_n(4000)>=1.4$) as defined earlier,
while the bottom panel shows the difference between the mean emission line component of SF galaxies and that of AGNs in their overlapping 
range of \myd4n\ ($1.3<D_n(4000)<1.6$). The spectral components were normalized to the total flux in the 16\mi\ IRS band. 
Among the most significant features are the decreased \neiii\ and \siii\ lines and increased H$_2S(0)$ line
in older SF galaxies with respect to younger ones, 
and the strong increase in \oiv\ line emission in AGNs along with diminished \neii\ and \siii\
emission with respect to SF galaxies. The H$_2$S(1) molecular line is also enhanced in AGNs. A strong excess of H$_2$ in many Seyferts
and LINERS has been reported by \cite{Roussel_etal2007}, suggesting a different excitation mechanism in these galaxies. H$_2$ line emission
is studied in more detail in Section 5.4.

While low excitation lines such as \neii\ and \neiii\ can be excited by hot stars as well as
AGNs (they are detected in all but 1 spectrum for \neii, all but 3 spectra for \neiii), 
the high excitation potential of the [OIV]25.89\mi\ line (54.9eV) (the brightest such line with [NeV]14.21\mi\
in the MIR) usually links it to AGN activity \citep[e.g.][]{Genzel_etal1998, Sturm_etal2002, Melendez_etal2008}.
However it has also been attributed to starburst related mechanisms \citep{SchaererStasinska1999, Lutz_etal1998}
and indeed detected in starburst galaxies or regions \citep{Lutz_etal1998, Beirao_etal2006, Alonso-Herrero_etal2009}.
It is detected in 73\% of our `pure' star-forming galaxies (63\% of the composite galaxies) while undetected in 1 out of 7 AGNs. 
It may also be that three quarters of our SF galaxies harbor an obscured AGN not detected in the optical. While the non detection 
of [OIV]25.89\mi\ in AGNs has also been known to happen \citep[e.g.][]{Weedman_etal2005}, the one AGN spectrum 
in our sample without  [OIV]25.89\mi\ (\#63)
is particularly noisy and the presence of the line, even significant, cannot be ruled out. 

\begin{figure}
\plotone{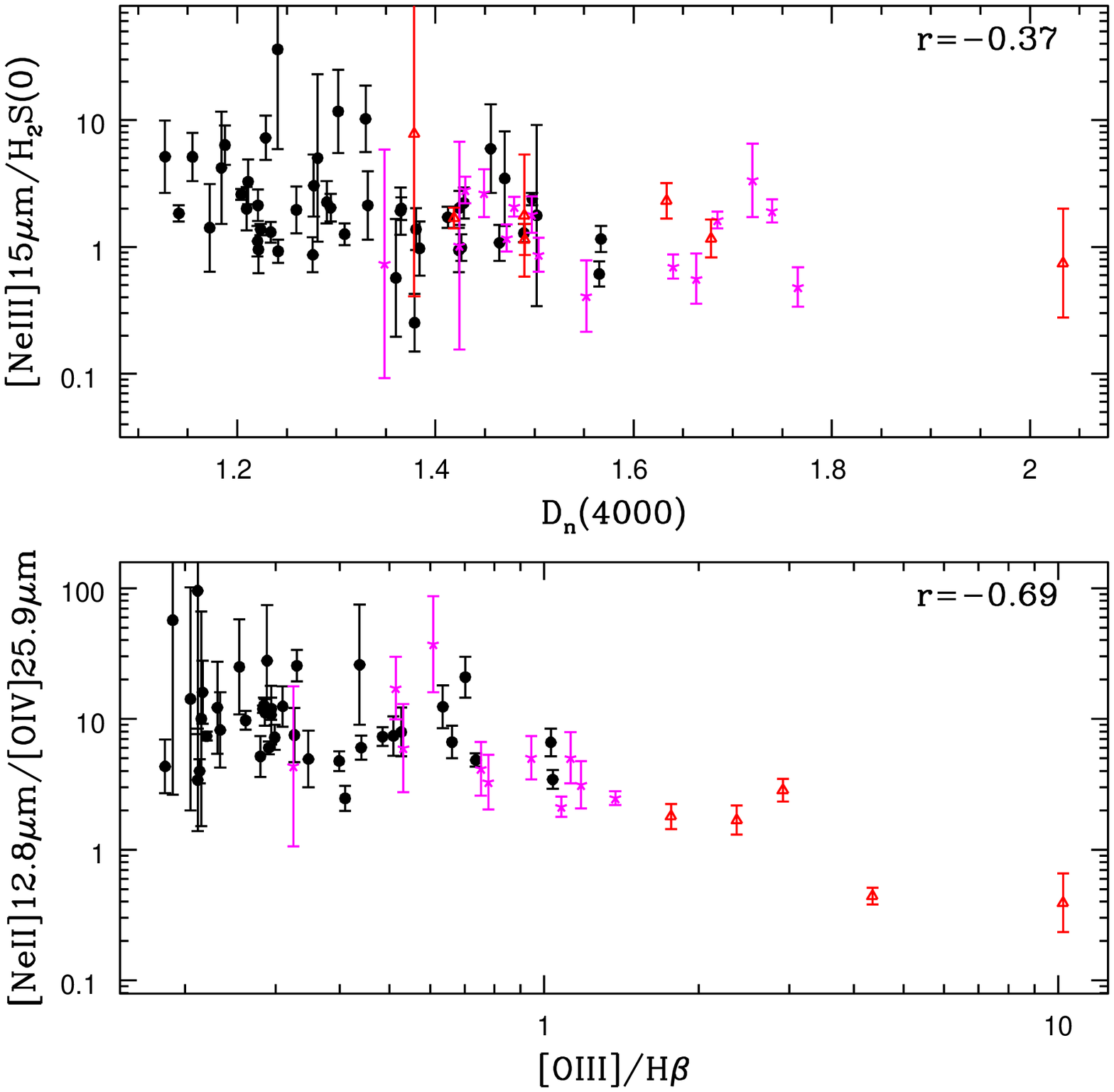}
\caption{ {\it Top:}  The \neiii/H$_2S(0)$ line ratios as a function of \myd4n.  
{\it Bottom:} The \neii/\oiv\ line ratios as a function of \myo3hb. Symbols are as described in Figure \ref{fig:cmd}. 
Ratios with extremely large errors are not shown. The pearson coefficients of the correlations are indicated in each panel
for the full sample. The most significant correlation is found between \neii/\oiv\ and \myo3hb\ for the subsample of 
composite galaxies and AGNs ($r=-0.80$). 
\label{fig:lineratios}}
\end{figure}

The top panel of Figure \ref{fig:lineratios} shows the \neiii/H$_2S(0)$ ratios as a function of \myd4n. 
The Pearson coefficient of the correlation is indicated in the top right corner. The trend is mild, 
and milder still for the \siii/H$_2S(0)$ ratios. Much more significant is the correlation between 
\neii/\oiv\  and \myo3hb\ shown in the bottom panel. The correlation for \siii/\oiv\ is somewhat
less significant but both ratios notably decrease with increasing radiation field hardness {\it for composite 
galaxies and AGNs} (the Pearson coefficient for this subsample is $r=-0.80$).
Ratios of high to low excitation emission lines have long been used to characterize the dominant source 
of ionization in galaxies \citep[e.g.][]{Genzel_etal1998}. We come back to this point in Section 3.3.

\begin{figure}
\plotone{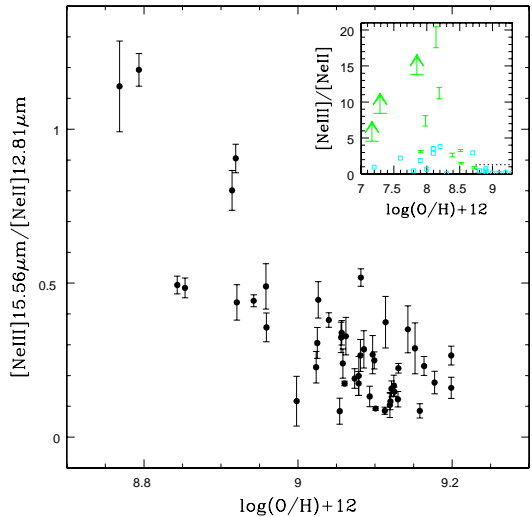}
\caption{The \neiii/\neii\ emission line ratios of star-forming galaxies as a function of metallicity. 
The inset shows the low metallicity data points of \cite{OHalloran_etal2006}  (open squares) and \cite{Wu_etal2006} 
(green error bars and lower limits), with our dynamic range shown as the dotted box in the bottom right corner.
\label{fig:Ne-Z} 
}
\end{figure}

The \neiii/\neii\ line ratio is also expected to be sensitive to the hardness of the radiation
field, however we find no correlation between this ratio and \myo3hb\ in our sample. We do find a trend with metallicity
despite the very narrow metallicity range of our sample, as shown in Figure \ref{fig:Ne-Z} for the SF subsample. 
Indeed \neii\ has been shown to be the dominant ionization species in \hii\ region at high metallicity while \neiii\ 
takes over in regions of lower density and higher excitation such as low mass, low metallicity galaxies \citep{OHalloran_etal2006, Wu_etal2006}. 
The inset shows a larger scale version of this figure with low metallicity data points from \cite{OHalloran_etal2006} 
(open squares) and \cite{Wu_etal2006} (green error bars and lower limits). Our dynamic range is represented as the 
dotted box in the bottom right corner.

\begin{figure}
\plotone{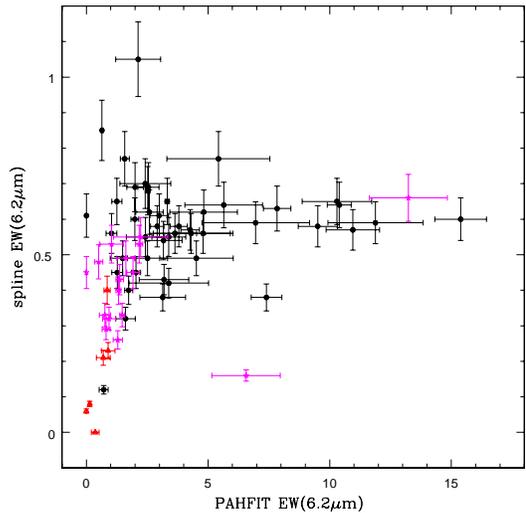}
\caption{The 6.2\mi\ PAH equivalent widths (EWs) computed using Eq. \ref{eq:ew} with the PAHFIT decomposition parameters compared
to the EWs computed by \cite{SargsyanWeedman2009} assuming a single gaussian on a linear continuum between 5.5\mi\ and 6.9\mi.
Symbols are as described in Figure \ref{fig:cmd}. 
\label{fig:ew-methods}
}
\end{figure}
\begin{figure*}[tb]
\plottwo{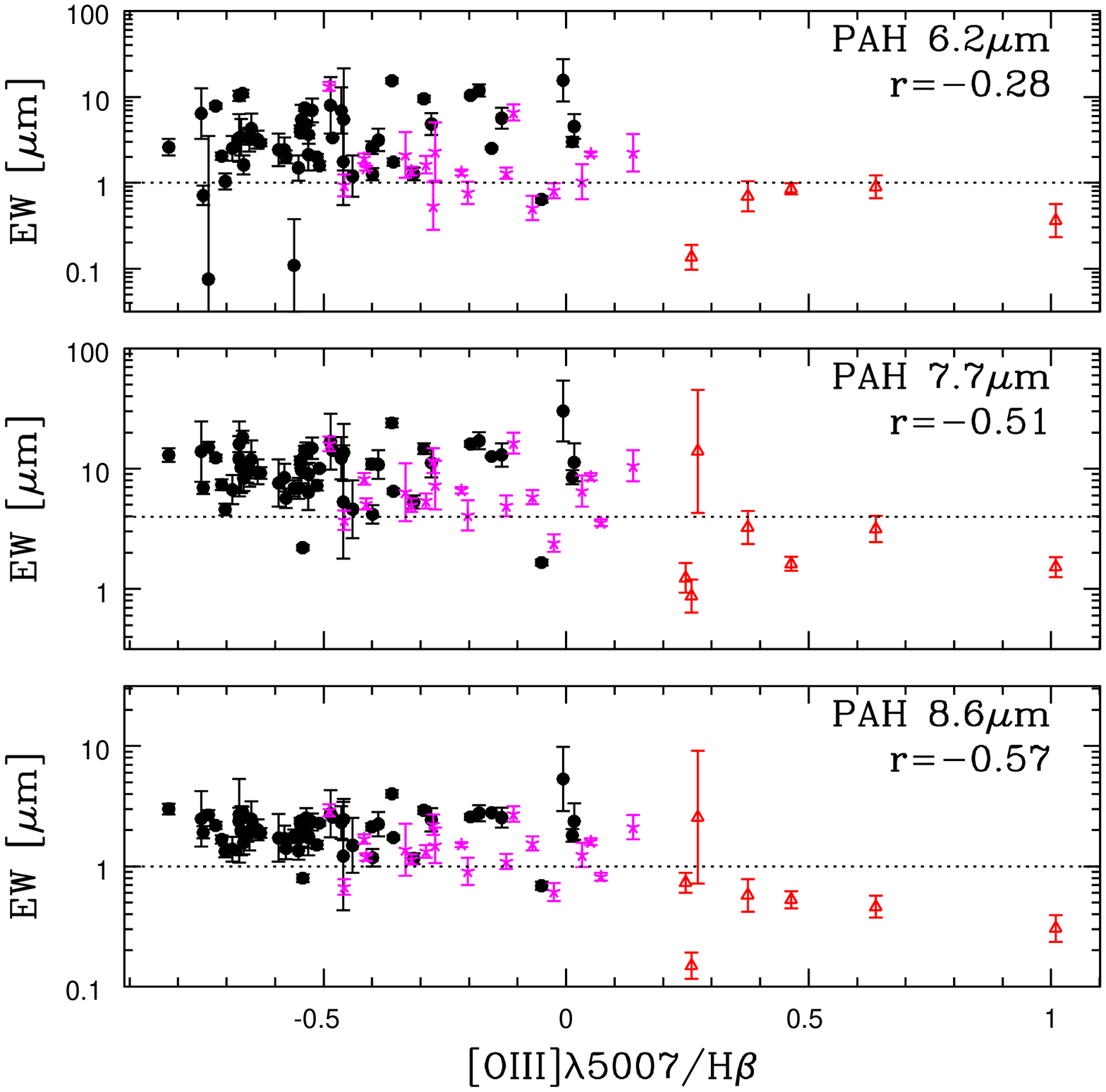}{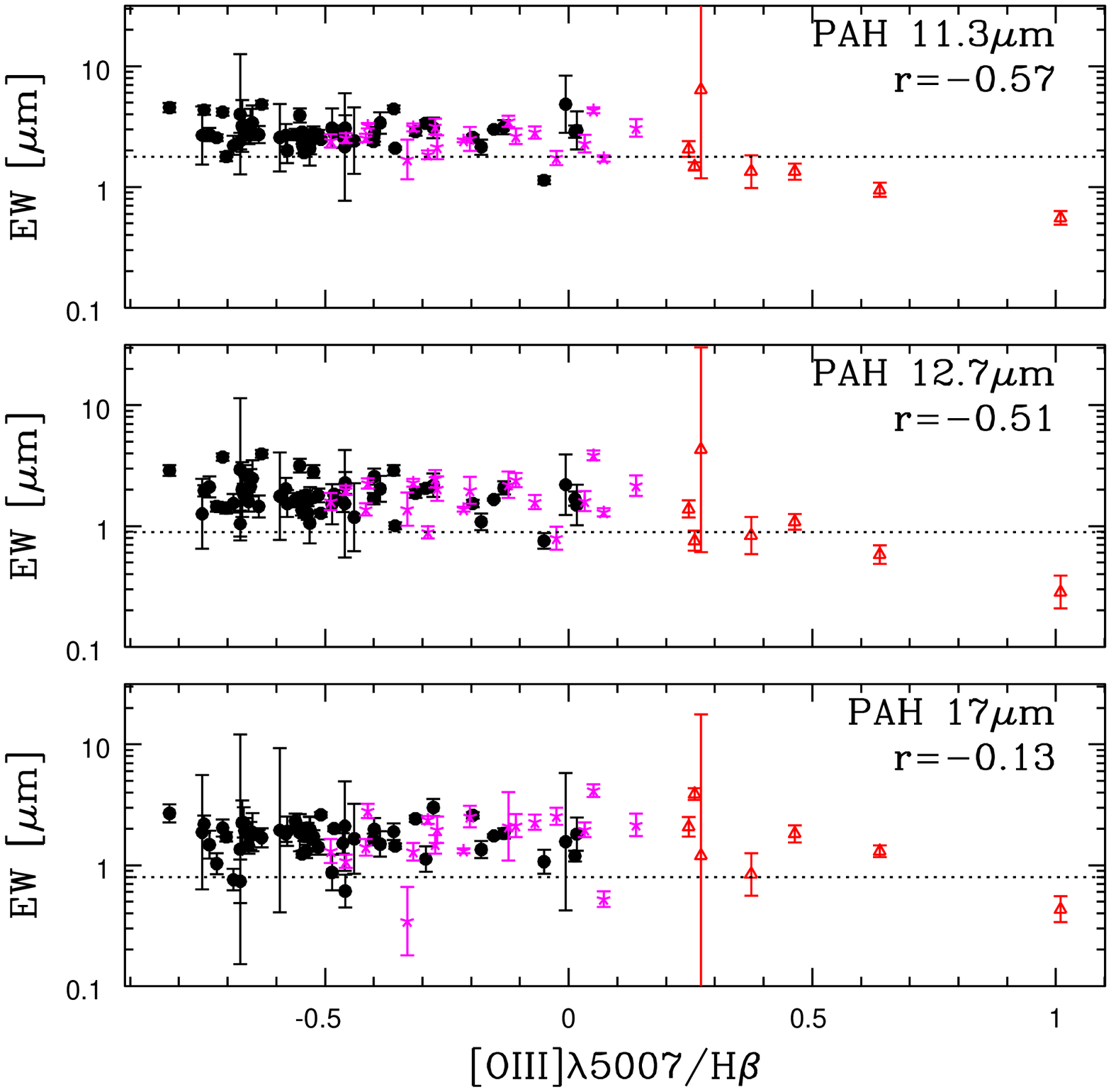}
\caption{The equivalent widths of the main PAH features at short wavelengths (6.2, 7.7 and 8.6\mi, left panel) 
and at long wavelengths (11.3, 12.7 and 17\mi, right panel) as a function of \myo3hb. 
Symbols are as described in Figure \ref{fig:cmd}. 
The Pearson correlation coefficients $r$ are indicated at the top right of each panel. 
The dotted lines in the left panel show the limits that best isolate SF galaxies (EW(6.2\mi) $>1 \mu m$, EW(7.7\mi) $>4\mu m$
and EW(8.6\mi) $>1\mu m$). The dotted lines in the right panel are approximate lower limits for the SF population
EW(11.3\mi) $>1.8 \mu m$, EW(12.7\mi) $>0.9\mu m$ and EW(17\mi) $>0.8\mu m$). AGN EWs become increasingly undistinguishable 
from those of SF galaxies towards longer wavelengths. 
\label{fig:ew-o3hb}
}
\end{figure*}

\subsection{PAH Equivalent Widths}

We compute equivalent widths (EWs) as the integrated intensity of the Drude profile(s) fitting a particular PAH feature,
divided by the continuum intensity below the peak of that feature. Using Eq. 3 from S07 for the integrated
intensity of a Drude profile, the EW of a PAH feature with central wavelength $\lambda_r$, full width at half-maximum $FWHM_r$ 
(as listed in S07, their Table 3), and central intensity $b_r$ (PAHFIT output), can be written as:
\begin{equation}
EW(\lambda_r)={\pi \over 2} { b_r \over I_{\nu}^{cont}(\lambda_r)}  ~{FWHM}_r 
\label{eq:ew}
\end{equation}
where $I_{\nu}^{cont}(\lambda_r)$ is the continuum component of Eq. \ref{eq:pahfit}. 
This definition is different from that of S07 in PAHFIT which computes the integral $ \int (I_{\nu}^{PAH}/I_{\nu}^{cont}) d\lambda$
in the range $\lambda_r \pm 6\times FWHM_r$. In the case of the 7.7 micron feature whose FWHM is large and extends 
the limit of the integral to regions beyond the IRS range where the continuum vanishes arbitrarily, the profile weighted 
average continuum is used. Despite this caveat, both methods agree within 10\% 
and the discrespancies 
virtually disappear when increasing the limits of the integral for all other PAHs\footnote{In the process of making these comparisons, 
we discovered two bugs in PAHFIT:  
1/ the code was mistakenly calling gaussian profiles instead of Drude profiles to compute the EW integral, thus underestimating
EWs by $\sim 1.4$, and 2/ it was applying silicate extinction to the continuum while using the extinction corrected PAH features
(according to Eq. 1, both components are equally affected by the extinction term).
These bugs are being corrected (J.D. Smith, private communication).}. 
However the EWs measured as above differ significantly from those estimated with the spline method, which consists in fitting a
spline function to the continuum from anchor points around the PAH feature, and a Gaussian profile to the continuum-subtracted feature. 
This method yields considerably smaller EW values as it assigns a non negligible fraction of the PAH flux extracted by PAHFIT to the continuum.
Figure \ref{fig:ew-methods} shows our EWs (Eq. \ref{eq:ew}) against the 6.2\mi\ PAH EWs computed by \cite{SargsyanWeedman2009} 
for the SSGSS sample assuming a single gaussian on a linear continuum between 5.5\mi\ and 6.9\mi. 
Their published sample is restricted to SF galaxies defined as having EW(6.2\mi) $>0.4$\mi\ \citep{WeedmanHouck2009}. The measurements
for the remaining galaxies were kindly provided by L. Sargsyan. Their formal uncertainty is estimated to be $\sim 10\%$. 
The two methods are obviously strongly divergent. The spline EWs strongly peak around a value of $\sim 0.6$\mi\ with no apparent 
correlation with the PAHFIT estimates, which reach $\sim 15\mu m$ and can be up to 25 times larger than the \cite{SargsyanWeedman2009} values.
Our EWs for the main PAH features are listed in Table \ref{table:eqw}.

The strength of a PAH feature depends on several interwined properties of the ISM: metallicity, radiation field
hardness, dust column density, size and ionization state distributions of the dust grains (Dale et al. 2006 \nocite{Dale_etal2006} and 
references therein). In particular it is shown to be reduced in extreme far-UV radiation fields, such as AGN-dominated environment
\citep{Genzel_etal1998, Sturm_etal2000, Weedman_etal2005}, near the sites of SF \citep{Geballe_etal1989, Cesarsky_etal1996, 
TacconiGarman_etal2005, Beirao_etal2006, Povich_etal2007,Gordon_etal2008} or in very low metallicity environments 
\citep{Dwek2005, Wu_etal2005, OHalloran_etal2006, Madden_etal2006}, where the PAH molecules are thought to get destroyed 
\citep[e.g.][]{Voit1992_pahdestruction}.

\begin{figure*}[tb]
\plottwo{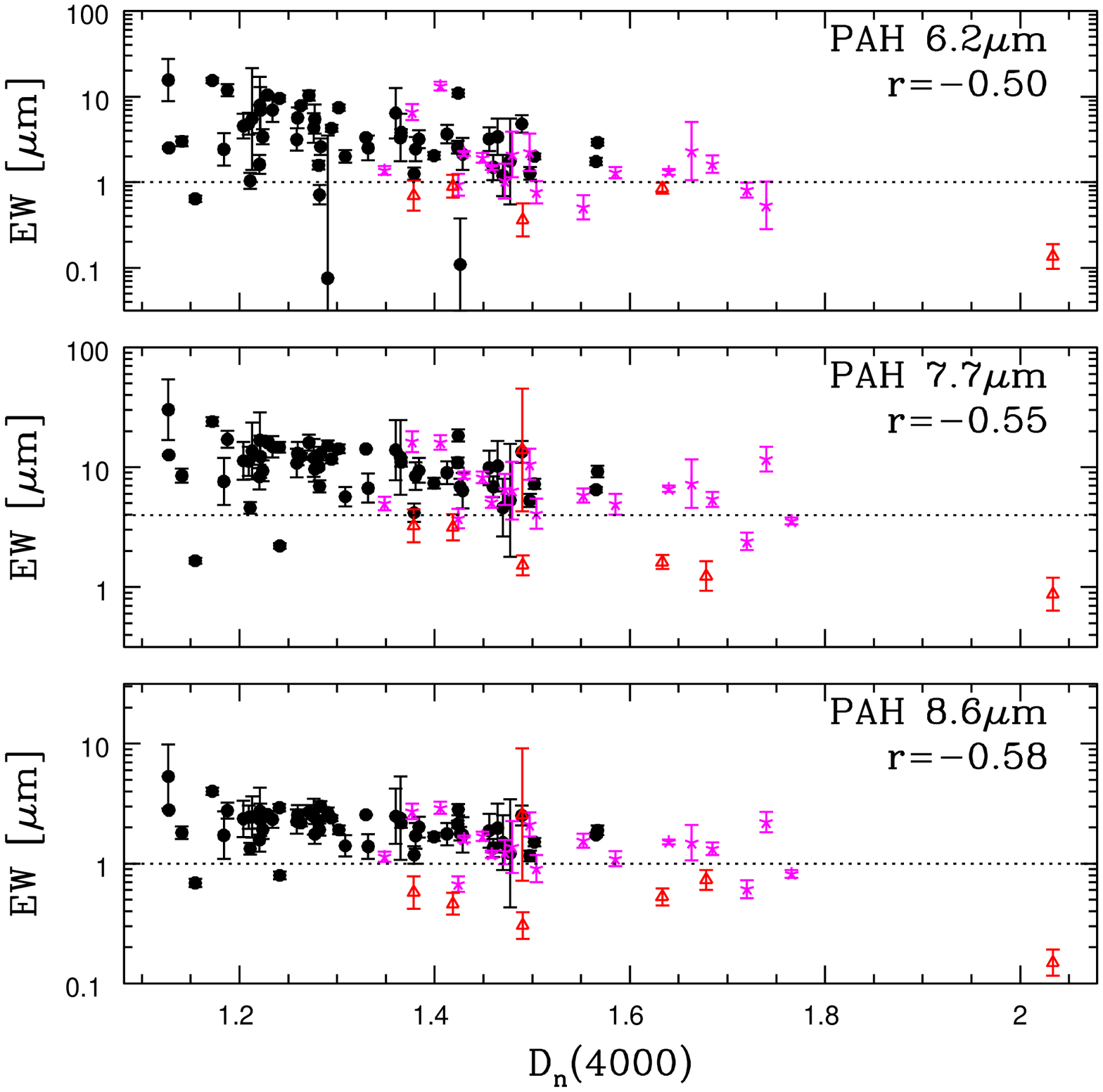}{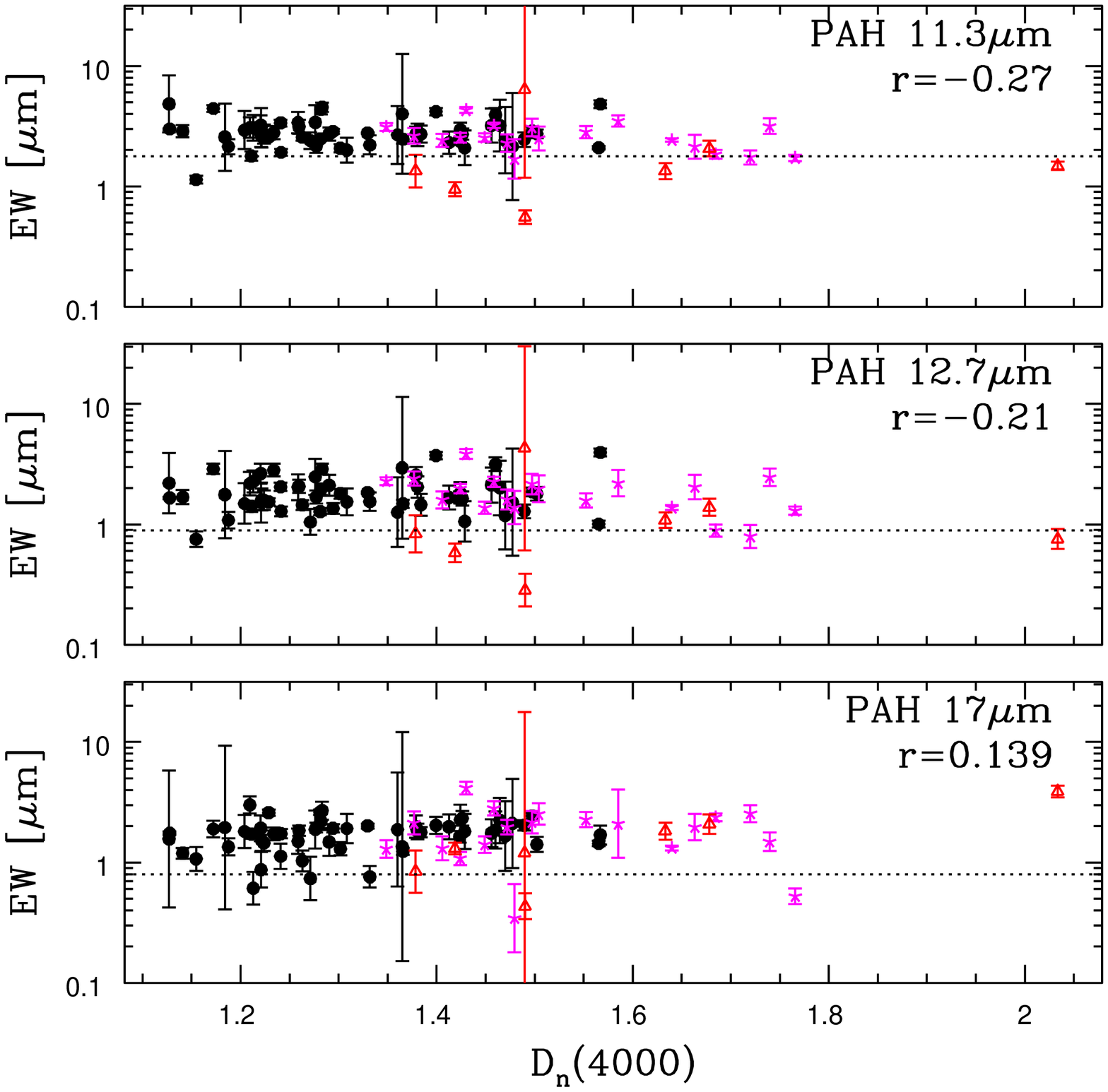}
\caption{The equivalent widths of the main PAH features at short and long wavelengths (left and right panels respectively)
as a function of \myd4n. Symbols and lines are as described in Figure \ref{fig:ew-o3hb}. 
\label{fig:ew-d4}
}
\end{figure*}
\begin{figure*}[tb]
\plottwo{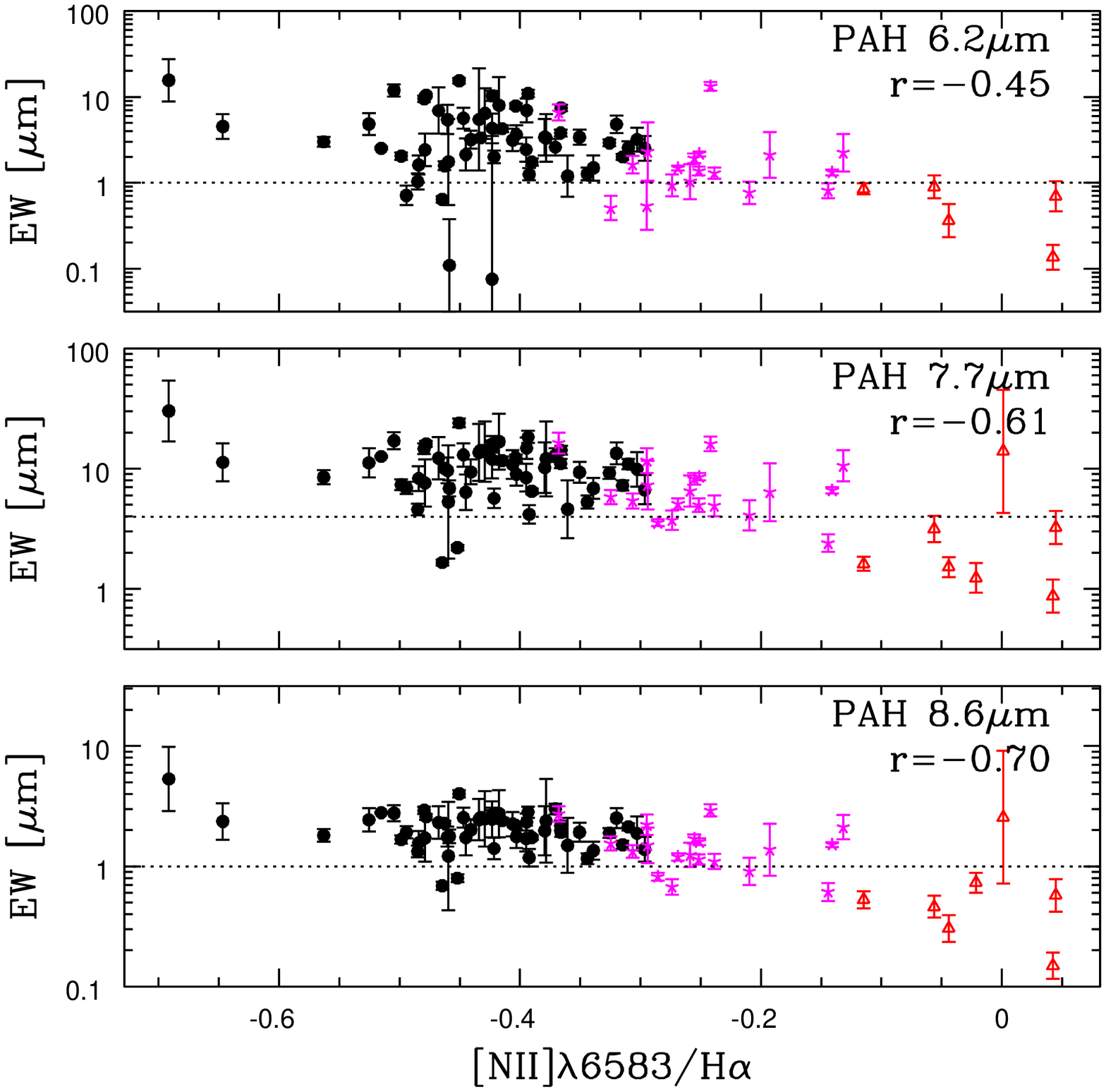}{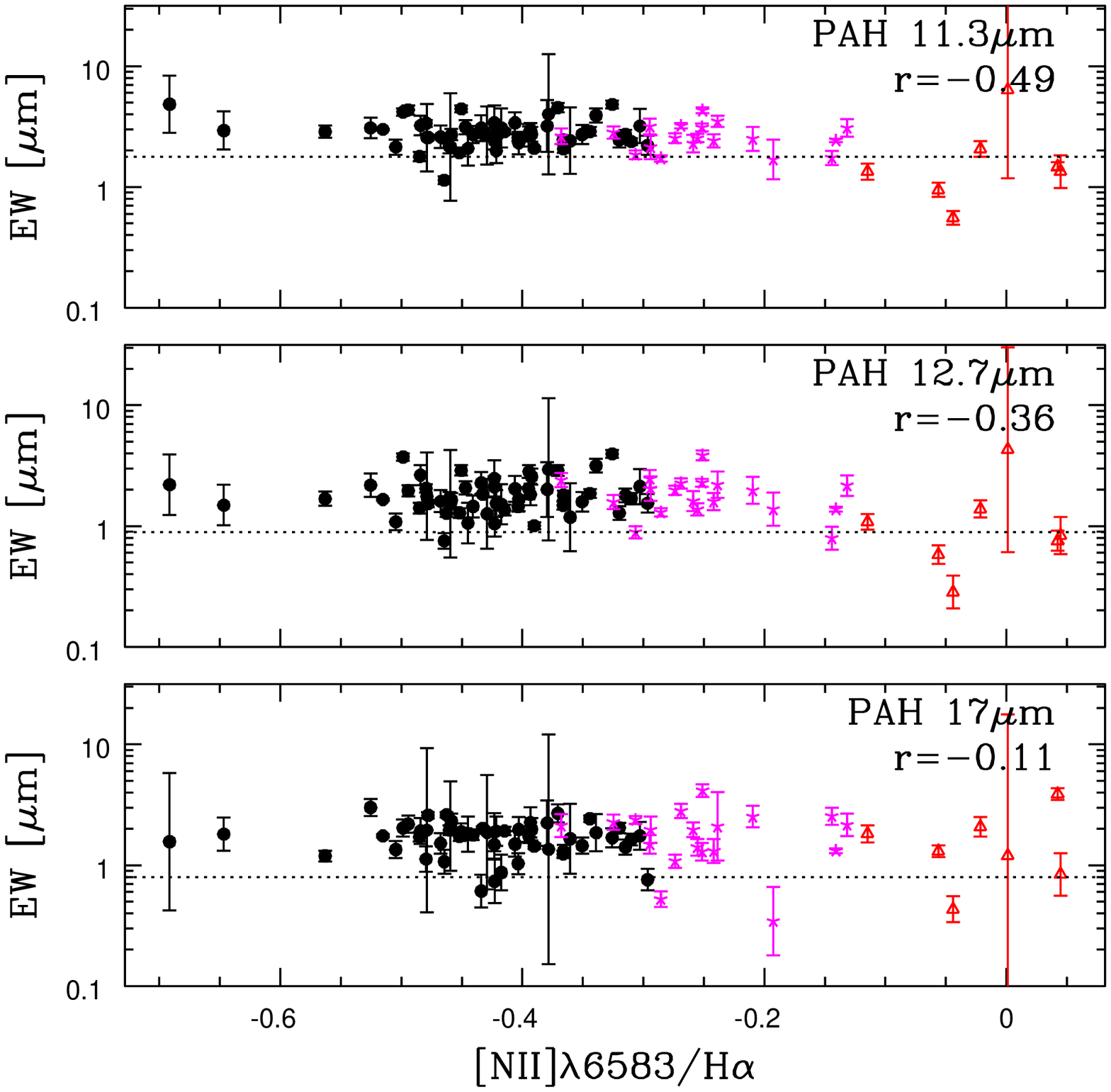}
\caption{The equivalent widths of the main PAH features at short and long wavelengths (left and right panels respectively)
as a function of \n2ha. Symbols and lines are as described in Figure \ref{fig:ew-o3hb}. The short wavelength PAH EWs 
significantly decrease with \n2ha.
\label{fig:ew-n2ha}
}
\end{figure*}

Figure \ref{fig:ew-o3hb} shows the EWs of the main PAH features as a function of \myo3hb.
The Pearson correlation coefficients $r$ are indicated at the top right of each panel. 
AGNs do exhibit noticeably smaller EWs than SF galaxies at short wavelengths  (6.2, 7.7 and 8.6\mi, left panel), 
however seemingly uncorrelated with radiation field hardness. 
The range of EWs spanned by AGNs becomes increasingly similar to that of SF galaxies towards longer wavelengths (11.3, 12.7 and 17\mi, right panel)
while at the same time a correlation seems to appear with radiation field hardness. The Pearson coefficients for the AGN population alone 
at long wavelengths are -0.97, -0.89 and -0.81 respectively from top to bottom, though admittedly they are boosted by the rightmost data point. 
A larger sample of AGNs is needed to confirm this correlation.
PAH strength remains largely independent of radiation field hardness for SF and composite galaxies. 
These results complement the analysis of \cite{ODowd_etal2009} who found a correlation between the long-to-short wavelength PAH ratios 
and \myo3hb in AGNs.
These trends are consistent with the selective destruction of PAH molecules in the hard radiation fields of these sources 
(\myo3hb $>1.5$). The EW trends or lack thereof in Figure \ref{fig:ew-o3hb} 
suggest that the smallest PAH molecules effective at producing the short-wavelength PAH features get destroyed first, 
near an AGN, while the larger molecules producing the larger wavelength PAHs require increasingly harder radiation fields for their 
PAH strength to drop below that of SF galaxies.  
\cite{DesertDennefeld1988} first suggested that the absence of PAHs could be taken as evidence for the presence of an AGN.
Weak PAH emission has since often been used to discriminate between photoionization and accretion disk processes. However the common boundaries  
for a `pure starburt', e.g. EW(7.7\mi)$>1$ \citep{Lutz_etal1998} or EW(6.2\mi) $>0.4$\mi\ \citep{WeedmanHouck2009} 
are significantly too weak here, due to the different method we use to compute the equivalent widths as shown above.
Based on the PAHFIT decomposition, SF galaxies would be best isolated by EW(6.2\mi) $>1$\mi, EW(7.7\mi) $>4$\mi\
or EW(8.6\mi)$>$1\mi, the latter two criteria being more accurately determined in our sample. Those limits are shown as
dotted lines in the left panel of Figure \ref{fig:ew-o3hb}. The two 
SF exceptions below the 7.7\mi\ and 8.6\mi\ EW limits 
(\#32 and \#74) happen to have very strong silicate absorption parameters ($\tau_{9.7}=$1.8 and 2.33) and still very distorted 
absorption-corrected continua compared to the rest of the sample. The dotted lines in the right panel are approximate lower limits
for the SF population (EW(11.3\mi) $>1.8 \mu m$, EW(12.7\mi) $>0.9\mu m$ and EW(17\mi) $>0.8\mu m$).
It is clear that the AGN population becomes increasingly difficult to isolate based on EW alone in
the red part of the spectrum.

Figures  \ref{fig:ew-d4} and \ref{fig:ew-n2ha} show the EWs of the main PAH features as a function \myd4n\ and 
\n2ha\ respectively. The EWs at short wavelengths show a mild downward trend 
with increasing age (or decreasing SF activities) while they become independent of it at long wavelengths.
This again is consistent with the correlations between 
the long-to-short wavelength PAH ratios and \myd4n\ or \ha\ equivalent width found by \cite{ODowd_etal2009}. 
The short wavelength EWs decrease more notably with increasing \n2ha\ ratios, which of course 
are related to \myd4n\ but appear to be the property that most uniformally and significantly
affects the sample as a whole. 
Metallicity and SF activity are known to affect PAH strength, however, as mentionned earlier, previous studies have demonstrated 
the opposite effect, namely a decrease in PAH strength at very low metallicity and in intense SF environment.
These trends thus make normal blue sequence galaxies the sites of maximum PAH strength.

\begin{figure}
\plotone{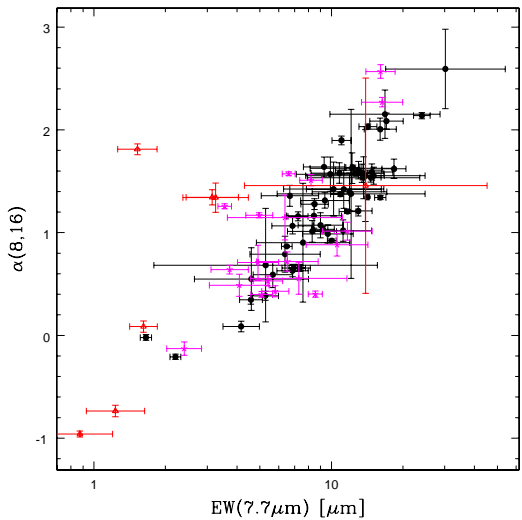}
\caption{The equivalent width of the 7.7\mi\ feature as a function of continuum slope (a diagnostic diagram proposed by \cite{Laurent_etal2000}).
Symbols are as described in Figure \ref{fig:cmd}. 
A clear trend is seen for SF and composite galaxies while AGNs appear more randomly distributed.
AGNs are expected to populate the lower left side of the plot, PDRs the lower right corner and \hii\ regions 
the upper left corner of the diagram \citep{Laurent_etal2000}.
\label{fig:laurent}}
\end{figure}

Other than PAH destruction, another cause of decreasing PAH strength at low wavelengths may be a stronger continuum
whose strength may depend on the above parameters.
A short wavelength continuum (3--10\mi) has been observed in AGNs, which is attributed to very hot dust heated by 
their intense radiation fields (Laurent et al. 2000 \nocite{Laurent_etal2000} and references therein), however the continuum slopes of AGNs
in our sample largely overlap those of the SF population (Figure \ref{fig:slope-x}). Figure \ref{fig:laurent} shows the
EW of the 7.7\mi\ feature as a function of continuum slope, a diagnostic diagram proposed by \cite{Laurent_etal2000} to distinguish
AGNs from PDRs and \hii\ regions. 
A clear trend is seen for SF and composite galaxies, suggesting that decreased PAH strength in normal SF galaxies may be  
at least partly due to an increased continuum at low wavelength, which is itself loosely correlated with \myd4n\ (Figure \ref{fig:slope-x}).
However the EWs of AGNs appear quite independent of their continuum slope, supporting the PAH destruction scenario.
In this diagram, AGNs are expected to populate the lower left side of the plot (shallow slopes and weak PAH features), PDRs the lower 
right corner (shallow slopes and strong PAH features) and \hii\ regions the upper left corner of the diagram (steep slopes and weak PAH features). 
Our SF sequence is qualitatively similar to the location of quiescent SF regions on the Laurent et al. diagram (their Figure 6), which 
are modeled by a mix of PDR and \hii\ region spectra, plus an AGN component towards the lower left corner where composite galaxies 
are indeed most concentrated.

\begin{figure*}[tb]
\plottwo{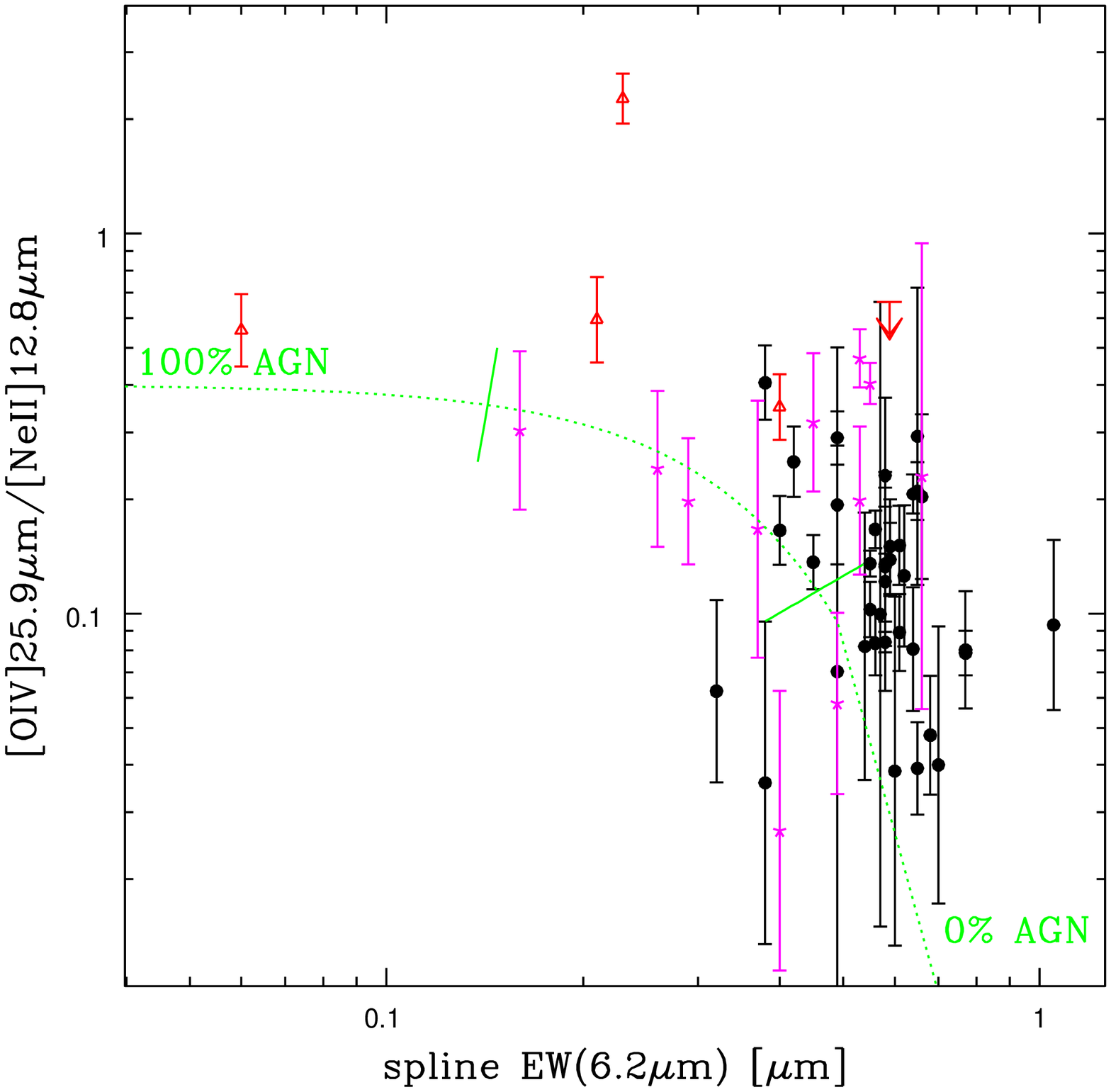}{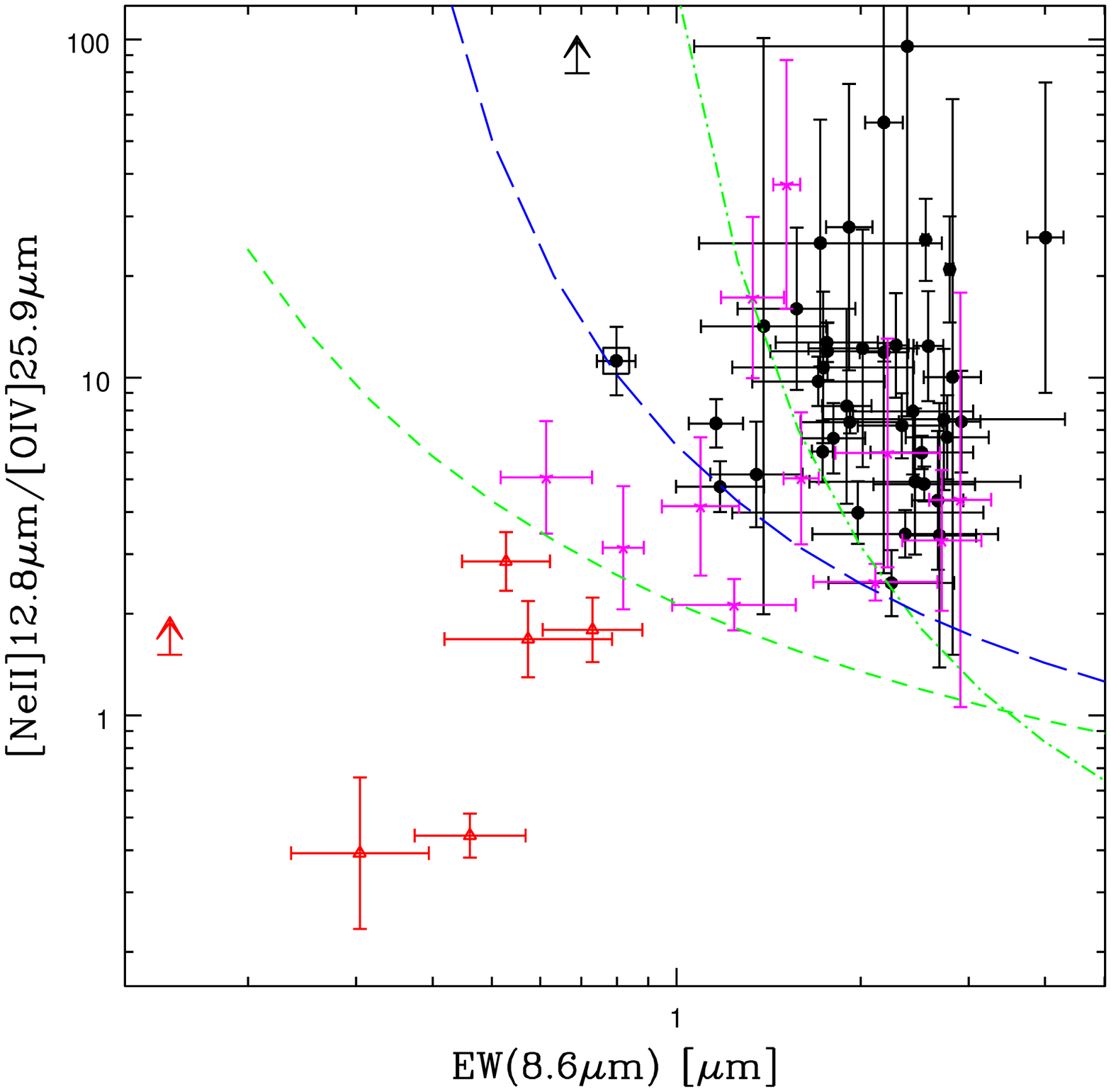}
\caption{{\it Left:} The spline derived 6.2\mi\ PAH equivalent widths \citep{SargsyanWeedman2009} against the \oiv/\neii\
cemission line ratios (a diagnostic diagram originally proposed by \cite{Genzel_etal1998}). 
The dotted line represents a variable mix of AGN and SF region; the short solid lines perpendicular
to it delineate the AGN region on the left, the SF region at the bottom right, and in between a region of mixed classifications
by \cite{Dale_etal2006}. 
We applied a cut in error bars for clarity. 
This diagram is of limited resolving power for normal galaxies. 
{\it Right:} The PAHFIT derived 8.6\mi\ PAH equivalent widths against \neii/\oiv\ (note the reversed
$y$-axis). This version which resembles a flipped version of the optical \cite{BPT1981} diagram better recovers the optical classification.
The short-dashed lower line and the dot-dashed upper line are optical boundaries translated into the MIR plane as explained in Section 3.3.
The long-dashed middle line is an empiral boundary marking the region below which we do not find any optically defined SF galaxy.
These boundaries are reported in Table \ref{table:diagnostics}.
The circled galaxy and the lower limit in the SF corner are the two SF galaxies with EW lower than the
SF limit in Figures \ref{fig:ew-o3hb}, \ref{fig:ew-d4} and \ref{fig:ew-n2ha} (EW(8.6\mi)$<$1\mi). 
\label{fig:mir-diagnostic}}
\end{figure*}

\subsection{Diagnostic Diagram} 

The presence of an AGN is thought to be best verified by the detection of strong high-ionization lines such as [NeV]14.21\mi\ or \oiv. 
\cite{Genzel_etal1998} were the first to show that the ratio of high to low excitation MIR emission lines combined with
PAH strength could be used to distinguish AGN activity from star-formation in ULIRGs. This diagnostic was recently revisited
by \cite{Dale_etal2006} for the nuclear and extra-nuclear regions of normal star-forming galaxies in the SINGS sample 
observed with the IRS. \cite{Dale_etal2006}
made use of the \oiv/\neii\ emission line ratios with spline derived EWs of the 6.2\mi\ feature  
(they also proposed an alternative diagnostic using the \si2/\neii\ emission line ratio but \si2\
is beyond the usable range of our data). 
The left panel of Figure \ref{fig:mir-diagnostic} shows the Dale et al. diagram using
the spline derived EWs of the 6.2\mi\ feature measured by \cite{SargsyanWeedman2009} in the SSGSS sample. 
The AGN with no detected \oiv\ line is plotted as an upper limit assuming an \oiv/\neii\
line ratio based on the correlation between \neii/\oiv\ and  \myo3hb\ for other AGNs in  
Figure \ref{fig:lineratios}. We applied a cut in error bars to this plot for clarity 
($\Delta$log(\neii/\oiv) $<1.5$, roughly the scale of the $y$-axis), which excludes 1 AGN, 1 composite galaxy
and 1 SF galaxy. One other AGN is found with no measurable EW. 
The dotted line represents a variable mix of AGN and SF region; the short solid lines perpendicular
to it delineate the AGN region on the left, the SF region at the bottom right, and in between a region of mixed classifications
whose physical meaning remains unclear \citep{Dale_etal2006}. Given the relative homogeneity of our sample (lacking in extreme types), 
the very narrow range of spline EWs for ordinary galaxies and the rather large uncertainties in our emission line ratios derived 
from low resolution spectra, this diagnostic proves of limited use for normal galaxies. Most optically classified SF galaxies 
do fall into the SF corner, but so do a few composite galaxies. The rest of the sample shows little spread within the mixed region.

Based on the results of this and the previous sections, we revise this diagnostic using the PAHFIT based EWs (Eq. \ref{eq:ew}) and the 
correlations between these EWs at low wavelength and \n2ha\ (Figure \ref{fig:ew-n2ha}) on the one hand, 
and between \neii/\oiv\ and  \myo3hb\ (Figure \ref{fig:lineratios}) on the other hand. 
The right panel of Figure \ref{fig:mir-diagnostic} shows the \neii/\oiv\ emission line ratios against the 
PAHFIT EWs of the 8.6\mi\ feature.
Note that we inverted the $y$-axis with respect to the left panel (and the traditional Genzel et al. diagram), so that 
the figure becomes a flipped version of the optical BPT diagram.
The short-dashed lower line is the theoretical optical boundary of \cite{Kewley_etal2001} translated into this plane using the correlations between 
EW(8.6\mi) and \n2ha\ in Figure \ref{fig:ew-n2ha} and between \neii/\oiv\ and  \myo3hb\
in Figure \ref{fig:lineratios} for the AGNs and composite galaxies. Its analytical form is:
\begin{equation}
y = { 1.84 \over x +1.51 } -0.88
\label{eq:mykewley}
\end{equation}
where $x=$ log(EW(8.6\mi)) and $y=$ log(\neii/\oiv). 
The dotted upper line is the empirical boundary of \cite{Kauffmann_etal2003} translated using these same correlations for the
composite and SF galaxies:
\begin{equation}
y = { 1.10 \over x +0.32 } -1.27
\label{eq:mykauffmann}
\end{equation}
As expected from the poorer correlation between \neii/\oiv\ and  \myo3hb\
for non AGNs, this boundary is less meaningful even though it does isolate the bulk of the SF galaxies.
The long-dashed  line is an empiral boundary marking the region below which we do not find any SF galaxy:
\begin{equation}
y = { 1.2 \over x+0.8 } -0.7
\label{eq:myboundary}
\end{equation}

Despite a mixed region of composite and SF galaxies, there is a clear sequence from the bottom left to the top right
of the plots and 3 regions where each optical class is uniquely represented. In particular weak AGNs separate remarkably 
well in this diagram. The mixed region may in fact be revealing an obscured AGN component in a large fraction ($\ge 50\%$) 
of the optically defined `pure' SF galaxies.  Other dust insensitive AGN diagnostics such as X-ray or radio data are necessary to confirm this. 
Deep XMM observations are available only over a small region of the Lockman Hole and the FIRST radio limits are
too bright to reliably test the presence of faint AGNs. Indeed we do not expect this hidden AGN contribution to be large since
none of the SF galaxies falls into the AGN corner of the diagram. These objects warrant a detailed study beyond the scope of 
the present paper.

Alternatively truly `pure' SF galaxies may be defined as lacking the \oiv\ emission line ($\sim 25\%$ of our SF
category). These are not represented except for one, which is one of the two SF galaxies with EWs lower than the
SF limit in Figs \ref{fig:ew-o3hb}, \ref{fig:ew-d4} and \ref{fig:ew-n2ha} (EW(8.6\mi)$<$1\mi). The lower limit was calculated by 
arbitrarily assigning it the lowest value of the [OIV] fluxes detected in the sample. The other one, which has a detected 
[OIV] line, is circled. These 2 galaxies which would have been misclassified as AGNs based on their EW alone sit well into 
the SF category on this diagram.
The AGN with no detected [OIV] (plotted as a lower limit) happens to have the lowest 8.6\mi\ EW in the sample. A significantly larger \neii/\oiv\ 
flux ratio would move it into the LINER region of this flipped BPT diagram (although this particular AGN is not optically classified as a LINER). 
Equations \ref{eq:mykewley}, \ref{eq:mykauffmann} and \ref{eq:myboundary} are reported in Table \ref{table:diagnostics} as well as their equivalents 
for the 6.2\mi\ and 7.7\mi\ PAH features.
We note that much larger samples, of AGNs in particular, are needed to confirm and/or adjust these relations.

\section{MIR dust components and the total infrared luminosity}

In this section we investigate how individual dust components emitting in the narrow MIR region 
trace the total dust emission in galaxies, which includes a very large FIR component. 
The definition of the total infrared (TIR) luminosity and the methods used to estimate it 
varies in the literature \citep{Takeuchi_etal2005}. In this paper \ltir\ refers to $L(3-1100\mu m)$ and has been 
derived by fitting the {\it Spitzer} photometric points (IRAC+IRS Blue Peak Up+MIPS) 
with \cite{DraineLi2007} model SEDs\footnote{http://www.astro.princeton.edu/$\sim$draine/dust/irem.html} and integrating 
the best fit SED from 3 to 1100\mi. 
This \ltir\ is in excellent agreement with the $3-1100$\mi\ luminosity 
derived from the prescription of \cite{DaleHelou2002} for MIPS data (their Eq. 4), with a standard deviation of 0.05 dex. This
shows that the total IR luminosity really is driven by the MIPS points \citep[e.g.][]{Dale_etal2007}.
We note also that integrating the SEDs between 8 and 1000\mi\ (sometimes called the FIR luminosity) would decrease the luminosity 
by $\sim 0.04$ dex in the present sample.

\begin{figure}
\plotone{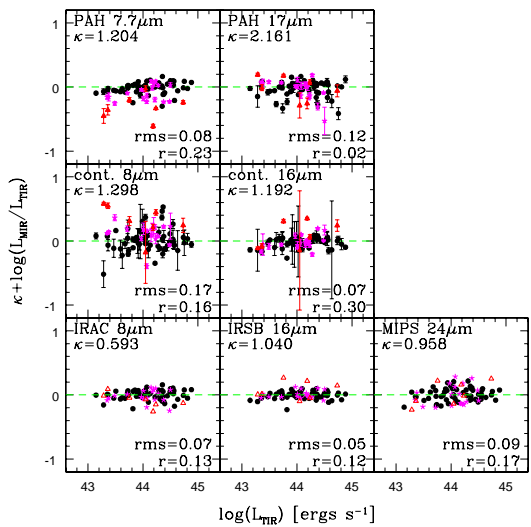}
\caption{$L_{MIR}/L_{TIR}$ ratios as a function of \ltir\ where $L_{MIR}$ equals - from top to bottom - 
the luminosity of the PAH complexes at 7.7 and 17\mi, the luminosity of the continuum at 8 and 16\mi, and 
the total restframe luminosities in the 8\mi\ IRAC band, 16\mi\ IRS band and 24\mi\ MIPS bands.
Symbols are as described in Figure \ref{fig:cmd}. 
The continuum and broadband luminosities are defined as $\nu L_{\nu}$.
The logarithmic scaling factors $\kappa$ indicated in each panel are defined as the median 
of log($L_{TIR}/L_{MIR}$) for the SF population alone (green dashed lines).
The rms and Pearson coefficient $r$ in each panel are also for the SF population alone.
\label{fig:ltir-all}
}
\end{figure}

Figure \ref{fig:ltir-all} shows the correlations between \ltir\ and $L_{MIR}/L_{TIR}$ ratios where $L_{MIR}$ equals - from top to bottom -
the luminosity of the PAH complexes at 7.7 and 17\mi, the luminosity of the continuum at 8 and 16\mi, 
and the total restframe luminosities in the 8\mi\ IRAC band, 16\mi\ IRS band and 24\mi\ MIPS bands. 
The continuum and broadband luminosities are defined as $\nu L_{\nu}$.
As in all previous figures, SF galaxies are shown as black dots, composite galaxies as pink stars and AGNs as red triangles.
The logarithmic scaling factors $\kappa$ indicated in each panel are defined as the median of log($L_{TIR}/L_{MIR}$) for the SF population
alone and is represented by the green dashed lines (${\rm log}(L_{MIR}/L_{TIR}) +\kappa =0$). 
The rms and Pearson coefficients of the correlations are also quoted for the SF population alone. 

It is striking that galaxies of all types follow the same tight, nearly linear correlations between \ltir\ and
the broadband luminosities in all 3 {\it Spitzer} bands over 2 dex in luminosity. This implies that all the galaxies in our 
sample are assigned nearly the same SED from a few \mi\ to a thousand \mi\ and that the FIR component can be well predicted from any 
one broadband luminosity in the MIR. This in turn suggests a common heating source for the small and large dust grains responsible for the
MIR and FIR emissions respectively \citep{Roussel_etal2001}. The same correlations apply whether this heating source is stellar or an AGN. 
Although this may result from the implicit stellar origin of the dust heating in the models, 
the source of ionizing radiation may not significantly affect the broad SED, at least for weak AGNs. 
Many attempts have been made to derive calibrations between \ltir\ and single MIR broadband luminosities \citep{CharyElbaz2001, 
Elbaz_etal2002, Takeuchi_etal2005, Sajina_etal2006, Brandl_etal2006, Bavouzet_etal2008, Zhu_etal2008}.
Our best fit slope at 16\mi\ (\ltir\ $\propto L_{16 \mu m}^{0.98\pm 0.02}$) is in good agreement 
with that of \cite{CharyElbaz2001} for the 15\mi\ ISO fluxes. At 24\mi\ our correlation for SF galaxies (\ltir\ $\propto L_{24 \mu m}^{0.94\pm 0.025}$) 
is more linear than found in other studies \citep{Takeuchi_etal2005, Sajina_etal2006, Zhu_etal2008, Bavouzet_etal2008} but the 
discrepancy with the first three calibrations \citep{Takeuchi_etal2005, Sajina_etal2006, Zhu_etal2008} disappears
when including composite galaxies into the fit (\ltir\ $\propto L_{24 \mu m}^{0.89 \pm 0.03}$). 
On the other hand our correlation is in excellent agreement with the \cite{MoustakasKennicutt2006} sample (hereafter MK06).

\begin{figure}
\plotone{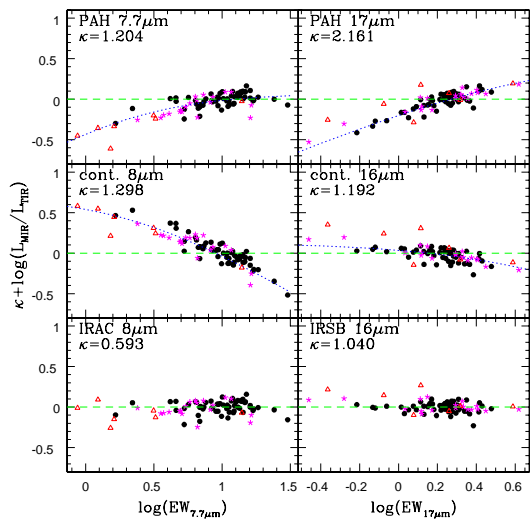}
\caption{$L_{MIR}/L_{TIR}$ ratios as a function of PAH equivalent width at 7.7\mi\ (left panels) and 17\mi\ (right panels)
where $L_{MIR}$ is defined at the top left of each panel. The symbols and $\kappa$ are defined as in Figure \ref{fig:ltir-all}. 
The dotted lines show the expected relations when the broadband fluxes at 8 and 16\mi\ are substituted for \ltir\ 
in the left and right panels respectively. 
\label{fig:residuals-ltir}
}
\end{figure}

The PAH and continuum luminosities also correlate remarkably tightly and nearly linearly with \ltir, however with some distinctions 
between AGNs and SF galaxies and between the hot and cool parts of the spectrum. The scatter between \ltir\ and  PAH luminosity for
SF galaxies is larger for the 17\mi\ PAH feature than for the 7.7\mi\ PAH feature. AGNs and composite galaxies blend with the SF 
population in the 17\mi\ feature correlation whereas they tend to have lower PAH luminosities at 7.7\mi\ and stronger 8\mi\ continua
for the same \ltir. The residuals are shown in Figure \ref{fig:residuals-ltir} 
as a function of the corresponding equivalent widths. The relations between these residuals and EWs are of course expected since 
the total flux at 8 and 16\mi\ can be nearly perfectly substituted for \ltir\ for SF galaxies and AGNs alike in the left and 
right panels respectively (the dotted lines show the predicted relations assuming these substitutions).
The most scattered correlation is found with the continuum luminosity at 8\mi. This may be due to larger measurement errors since
this continuum is faint and/or a stellar contribution unrelated to \ltir. A more speculative reason may be that this continuum  
originates from high intensity radiation fields only and is thus uncorrelated with the cold component of \ltir, unlike the PAH
emission.

The scaling factors $\kappa$ 
are listed in Table \ref{table:coeffs} for the main PAH features and the {\it Spitzer} band luminosities.
We also add to our list of MIR components the peak luminosity of the 7.7\mi\ PAH feature, defined as $\nu L_{\nu}(7.7\mu m)$, 
as it is a more easily measurable quantity at high redshift than the integrated flux of the PAH feature \citep{WeedmanHouck2009, SargsyanWeedman2009}.
For galaxies with EW$>4$\mi\ (most SF galaxies), the median ratio of this peak luminosity to 
the total luminosity of the PAH complex, $\nu L_{\nu}(7.7\mu m)/L_{PAH}(7.7\mu m)$, is $9.3 \pm 0.9$ and the peak luminosity 
estimates the total PAH luminosity to within $\sim$ 20\%. However the overestimate can be as large as 50\% for other galaxy types in this
sample, in particular galaxies containing an AGN which may not be easily isolated in high redshift samples and may also have much smaller 
EWs leading to yet larger errors.

For the calibration that shows the strongest deviation from linearity in Figure \ref{fig:ltir-all}, 
which is found for the 7.7\mi\ PAH luminosity ($L_{TIR}\propto L_{MIR}^{0.93\pm 0.02}$),
the linear approximation ${\rm log}(L_{TIR})= {\rm log}(L_{PAH}(7.7\mu m))+\kappa$ (where $\kappa=1.204$) recovers \ltir\ within a factor 
of 1.5 {\it in this sample}. 
For starbursts and ULIRG starbursts, \cite{Rigopoulou_etal1999} found a mean log$(L_{TIR}/L_{PAH}(7.7\mu m))$ of 2.09 and 2.26 respectively,
considerably larger than for normal galaxies. More recently \cite{Lutz_etal2003} find a mean logarythmic ratio of 1.52 for 
a sample of starburst nuclei, closer to our value.
Our mean logarythmic ratio for the 6.2\mi\ feature is 1.5 and 2.0 with and without aperture correction respectively 
while \cite{Spoon_etal2004} find a value of 2.4 for a sample of normal and starburst nuclei. This ratio is yet higher (3.2) in Galactic \hii\ regions 
while highly embedded star-forming regions can lack PAH emission altogether \citep{Peeters_etal2004}. 
These increased ratios for starburst regions compared to normal SF galaxies are generally attributed to PAH destruction near the site of on-going SF
due to intense radiation fields, making PAHs poor tracers of SF (Peeters et al. 2004 and references therein). 
The EW dependence of the log$(L_{TIR}/L_{PAH}(7.7\mu m))$ ratio is clearly seen within our sample in the upper 
left panel of Figure \ref{fig:residuals-ltir}. This cautions against the use of a single linear relation between PAH luminosity and \ltir\ 
for galaxies of unknown physical properties. 

However independently of galaxy type we expect to find lower values than these studies which all made use of interpolation methods to extract
the PAH features. Using a Lorentzian profile fitting method comparable to PAHFIT for a sample of starburst-dominated LIRGS at $z\sim 0.5-3$, 
\cite{HernanCaballero_etal2009}
find mean log$(L_{TIR}/L_{PAH})$ ratios of $1.92\pm 0.25$, $1.42\pm 0.18$ and $1.96\pm 0.27$ for the 6.2, 7.7 and 11.3\mi\ features respectively.
These ratios are 
2.6, 1.8, and 1.4 times larger than ours respectively,
closer than previous studies despite the quite different galaxy type considered. The wavelength gradient 
can be explained in the context of selective PAH destruction.

Finally we note that in our sample the total $6.2-33$\mi\ PAH luminosity amounts 
to $\sim 15\%$ of the total IR luminosity for SF galaxies, $\sim 11\%$ for composite galaxies and $\sim 8\%$ for AGNs.
The 7.7\mi\ feature alone accounts for $\sim 40\%$ of the total PAH emission. These fractional contributions
are in good agreement with those found in the SINGS sample (S07).

\begin{figure}
\plotone{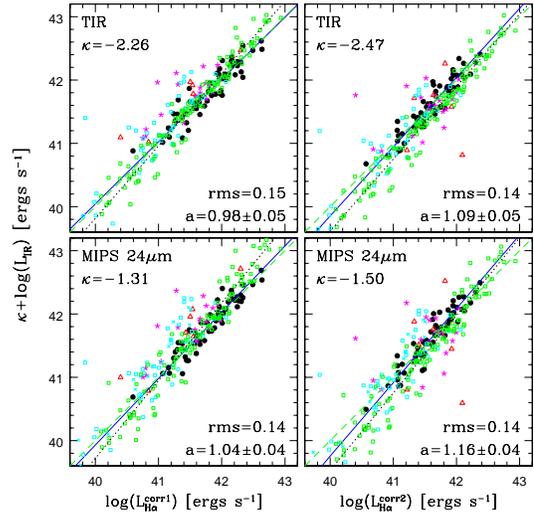} 
\caption{{\it Left:} the extinction and $r$-band aperture corrected \ha\ luminosity against the TIR and 24\mi\ continuum
luminosities. $\kappa$ is defined as the median of log($L_{{\rm H}\alpha}^{corr}/L_{IR}$) for the SF population alone. 
The rms and slope of the linear regressions (solid lines) are also shown for the SF population. The green dashed lines indicate 
equality. The open blue squares and crosses are SINGS data (integrated values and galaxies centers respectively); The open green squares 
represent the \cite{MoustakasKennicutt2006} (MK06) sample. The dotted lines are fits to the MK06 sample. 
{\it Right:} same as in the left panels but using the smaller \ha\ aperture corrections computed by \cite{B04} (see text for details). 
The new correlations (solid lines) are steeper, in better agreement with data that do not require aperture corrections.
\label{fig:lha-ltir}}
\end{figure}

\section{MIR components and the Star-Formation Rate}

The TIR luminosity is a robust tracer of the SFR for very dusty starbursts, whose stellar emission is dominated by 
young massive stars and almost entirely absorbed by dust (typically galaxies with depleted PAH emission),
but for more quiescent and/or less dusty galaxies such as those in the present sample, it can include a non negligible 
contribution from dust heated by evolved stars (`cirrus emission') as well as miss a non negligible fraction of the young stars' 
emission that is not absorbed by dust \citep{LonsdaleHelou1987}.
For normal spiral galaxies the contribution of non ionizing photons may actually dominate the dust heating over \hii\ regions 
\citep{Dwek_etal2000, Dwek2005} while low dust opacity makes these galaxies \ha\ and UV bright.
The tight correlations between MIR luminosities and \ltir\ indicate that the same caveats apply from the MIR to the FIR
\citep{Boselli_etal2004}.

\subsection{MIR dust and \ha}

\ha\ emission is a more direct quantifier of young massive stars - in the absence of AGN -
but inversely it must be corrected for the fraction that gets absorbed by dust. 
The SDSS line fluxes are corrected for foreground (galactic) reddening using \cite{ODonnell1994}.
The correction for intrinsic extinction is usually done using the Balmer decrement and an extinction curve to first order, 
or more accurately with higher order hydrogen lines \citep{B04}. 
Here we correct the SDSS \ha\ fluxes in the usual simple way using the stellar-absorption
corrected \ha/\hb\ ratio and a Galactic extinction curve. We assumed an intrinsic \ha/\hb\ ratio of 2.86
(case B recombination at electron temperature $T_e=10 000K$ and density $N_e=100~{\rm cm}^{-3}$) and
$R_V=A(V)/E(B-V)=3.1$ (the mean value for the diffuse ISM). 
The \ha\ attenuations range from 0.4 to 2.3 mag in the SF galaxy subsample with a median value of 1.1 mag, 
meaning that between 10 and 70\% of the \ha\ photons do {\it not} get reemitted in the IR.

The SDSS \ha\ measurements also require fiber aperture corrections. 
Here again we apply the usual method which consists in scaling the fiber-measured \ha\ fluxes using the $r$-band Petrosian-to-fiber 
flux ratios \citep{Hopkins_etal2003}. The mean value for these ratios is 3.5. 
The left panels of Figure \ref{fig:lha-ltir} shows the extinction and aperture corrected \ha\ luminosity, $L_{{\rm H}\alpha}^{corr}$, 
against the TIR and 24\mi\ continuum luminosities (top and bottom panel respectively). 
The rms and slope $a$ of the linear regressions (solid lines) are indicated for the SF population alone. 
The logarithmic scaling factors $\kappa$ indicated in each panel are defined as the median of log($L_{{\rm H}\alpha}^{corr}/L_{IR}$) also
for the SF population alone. The green dashed lines indicate equality (${\rm log}(L_{{\rm H}\alpha}^{corr})={\rm log}(L_{IR}) +\kappa$). 
Overlaid are the MK06 data (open green squares) and SINGS data (open blue squares for the integrated measurements, 
crosses for 20''x20'' galaxy center measurements), taken from \cite{Kennicutt_etal2009} (hereafter K09). 

Our median \ltir\ to  $L_{{\rm H}\alpha}^{corr}$ logarithmic ratio of $2.27 \pm 0.2$ 
is in good agreement with the ratio of SFR calibration coefficients derived by \cite{Ken98}  for \ha\ and \ltir\ respectively,
implying that \ltir\ (and the MIR components that correlate with it) may be reasonable SFR tracers in normal SF galaxies 
after all. This may actually be a coincidence due to the fact that the cirrus emission and the unattenuated ionizing flux
roughly cancel each other in massive spiral galaxies (K09 and references therein).
Our $L_{TIR}/L_{{\rm H}\alpha}^{corr}$ ratio is also in good agreement with the MK06 sample ($2.32\pm 0.19$). 
In recent years several groups have exploited the capabilities of {\it Spitzer} to re-investigate the relationship 
between MIR components and \ha\ emission. 
Our mean $\nu L_\nu(24\mu m)$ to  $L_{{\rm H}\alpha}^{corr}$ logarithmic ratio of $1.31 \pm 0.14$ is in good agreememt with 
these \citep[e.g.][]{Wu_etal2005,Zhu_etal2008,Kennicutt_etal2009}, 
as is the higher mean $\nu L_\nu(24\mu m)/L_{{\rm H}\alpha}^{corr}$ 
ratio for composite galaxies \citep{Zhu_etal2008}. However the slope of our correlations for SF galaxies tend to be more linear 
than found in these studies (the dotted lines in Figure \ref{fig:lha-ltir} show fits to the MK06 sample).
Yet non linearity is expected from the positive correlation between attenuation and SFR. 
Given the good agreement between our and the MK06 samples in the IR (cf. the \ltir\ --  $\nu L_\nu(24\mu m)$ correlation
in the previous section), differences in \ha\ measurements must be responsible for the discrepancy in slopes. In particular it is possible 
that aperture corrections, which are not needed for the MK06 sample, are overestimated for all or a fraction of our galaxies. This would
be the case if SF is more intense at the center of the galaxies and/or more attenuated, a common occurence in spiral galaxies
\citep[e.g.][]{Calzetti_etal2005}. 
 
As a test we consider the smaller aperture corrections derived by \cite{B04} (hereafter B04)
that rely on the likelihood distribution of the specific SFR as a function of colors. These corrections
depend on the galaxy colors outside the fiber which are not necessarily the same as inside, and are on average
$\sim 1.6$ smaller than the $r$-band corrections for SF galaxies.
The right panels of Figure \ref{fig:lha-ltir} shows the same relations as in the left panels using these smaller aperture 
corrections. The new correlations (solid lines) are indeed steeper while the higher mean $L_{TIR}/L_{{\rm H}\alpha}^{corr}$ and 
$\nu L_\nu(24\mu m)/L_{{\rm H}\alpha}^{corr}$ logarithmic ratios of $2.47\pm 0.14 $ and $1.50\pm 0.15$ respectively 
remain within the range of the MK06 sample. 

\begin{figure}
\plotone{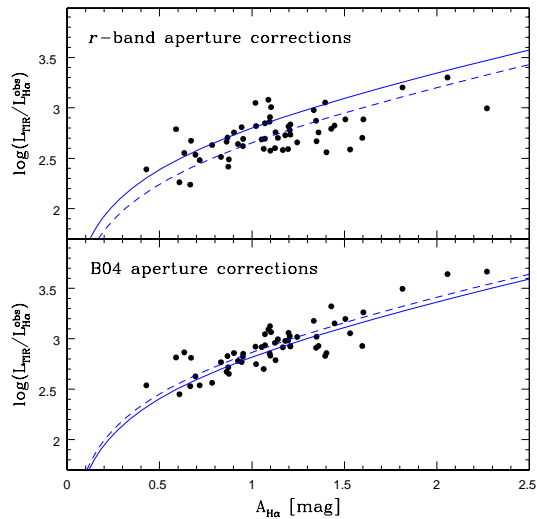}
\caption{The ratio of \ltir\ to observed \ha\ luminosity as a function of \ha\ attenuation measured from the Balmer decrement
(SF galaxies only).
The top panel assumes conventional $r$-band aperture corrections for \ha, while the bottom panel assumes the B04
aperture corrections (see text for details). The solid lines are best fits to Eq. \ref{eq:Aha} by K09 
for the SINGS+MK06 samples ($a_{TIR}=0.0024 \pm 0006$).
The dashed lines are best fits to the SSGSS sample ($a_{TIR}=0.0033 \pm 0.0014$ in the top panel, $0.0020 \pm 0.0006$ in the bottom panel). 
The smaller aperture corrections used in the bottom panel significantly improves the fit and the agreement between the three samples.
\label{fig:tir-Aha}
}
\end{figure}

More dramatic is the effect on the relation between \ha\ attenuation and the ratio of \ltir, or other IR luminosity,
to $L_{{\rm H}\alpha}^{obs}$, the `observed' (aperture-corrected but attenuation-uncorrected) \ha\ luminosity.
This relation is shown in Figure \ref{fig:tir-Aha} for both types of aperture correction. 
K09 modelled the \ha\ attenuation as:  
\begin{equation}
A_{{\rm H}\alpha}=2.5~{\rm log}\left[1+ a_{IR} {L_{IR}\over L_{{\rm H}\alpha}^{obs}}\right],
\label{eq:Aha}
\end{equation}
equivalent to  $L_{{\rm H}\alpha}^{corr}=L_{{\rm H}\alpha}^{obs}+ a_{IR} L_{IR}$. This energy balance approach 
was introduced by \cite{Calzetti_etal2007}, \cite{Prescott_etal2007} and \cite{Kennicutt_etal2007} to correct \ha\ fluxes
but has long been used to estimate UV attenuations from the $L_{TIR}/L_{FUV}$ ratios \citep[e.g.][]{Meurer99}.
The solid lines in both panels of Figure \ref{fig:tir-Aha} show the best fits by K09 for the SINGS and MK06 samples
($a_{TIR}=0.0024 \pm 0.006 $). The dashed lines are best fits for the SSGSS sample ($a_{TIR}=0.0033 \pm 0.0014$ in the top panel
and $0.0020 \pm 0.0006$ in the bottom panel). The smaller aperture corrections used in the bottom panel significantly improves 
the fit and the agreement between the three samples. Unless otherwise stated we now assume these corrections in the rest of the
paper. 

\begin{figure}
\plotone{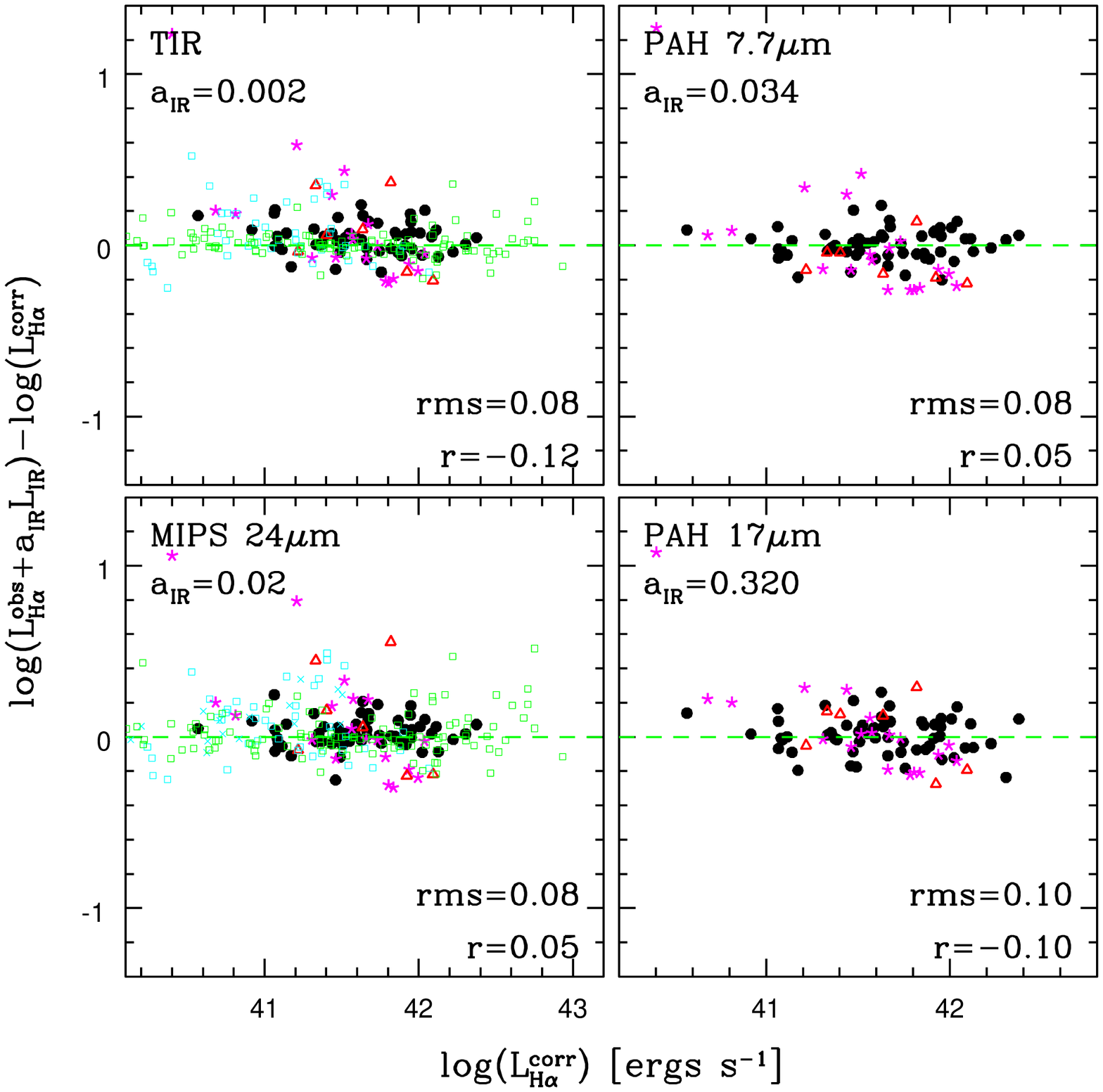}
\caption{$L_{{\rm H}\alpha}^{obs}+a_{IR} L_{IR}$ to  $L_{{\rm H}\alpha}^{corr}$ ratios as a function of $L_{{\rm H}\alpha}^{corr}$
for the TIR, 24\mi\ continuum, 7.7\mi\ and 17\mi\ PAH luminosities, assuming the B04 aperture corrections for \ha. 
The $a_{IR}$ coefficients are indicated at the top left of each panel. For the TIR and 24\mi\ luminosities, 
$a_{TIR}=0.0024$ and $a_{24}=0.020$ are best fits to Eq. \ref{eq:Aha} for the SINGS+MK07 samples by K09.
For the PAH lumosities, $a_{7.7\mu m}=0.034 \pm 0.012$ and $a_{17\mu m}=0.320 \pm 0.159$ are best fits to Eq. \ref{eq:Aha} for 
the SSGSS sample. 
\label{fig:lha-sample-ken}
}
\end{figure}

Figure \ref{fig:lha-sample-ken} shows the $L_{{\rm H}\alpha}^{obs}+ a_{IR}L_{IR}$ to $L_{{\rm H}\alpha}^{corr}$ ratios as a function
of  $L_{{\rm H}\alpha}^{corr}$ for the TIR, 24\mi\ continuum, 7.7\mi\ and 17\mi\ PAH luminosities.
The $a_{IR}$ coefficients are indicated at the top left of each panel for the SF population. The rms and Pearson coefficients are also 
indicated for the SF population.
For the TIR and 24\mi\ luminosities, $a_{TIR}=0.0024\pm 0.0006$ and $a_{24}=0.020\pm 0.005$ are best fits to Eq. \ref{eq:Aha} 
for the SINGS+MK06 samples by K09.  The combinations of \lhaobs\ and \ltir\ or $\nu L_\nu(24\mu m)$ provide a very
tight (rms=0.08) and perfectly linear fit to the total \ha\ luminosity
for all samples combined over 5 dex in luminosity, as was  shown by K09 for
the SINGS and MK06 samples. Composite galaxies follow nearly the same
relation save for 2 over-corrected outliers. Although more
scattered AGNs also follow the SF population.  For the PAH 
luminosities, $a_{7.7\mu m}=0.034 \pm 0.012$ and $a_{17\mu m}=0.320 \pm 0.159$ are best fits to Eq. \ref{eq:Aha} for the present sample.  
Here also the combinations of $L_{{\rm H}\alpha}^{obs}$ and PAH luminosities provide a much improved fit to the total \ha\ luminosity
compared to the raw $L_{PAH}/L_{{\rm H}\alpha}^{corr}$ relations (not shown), including for composite galaxies and AGNs
with the exception of a few outliers, most notably a composite galaxy with no \ha\ attenuation and a large IR/\ha\ ratio (\#84).

The same exercise can be performed with similarly good results with any other MIR dust components. The $a_{TIR}$ and $a_{MIR}$ coefficients 
for the SSGSS sample are listed in Table \ref{table:coeffs}. Note that $a_{MIR} \sim 10^{\kappa}a_{TIR}$, using the scaling factors 
$\kappa=<{\rm log} (L_{TIR}/L_{MIR})>$ listed in the first column 
of Table \ref{table:coeffs}. Although the $\kappa$ factors and $a_{TIR}$ depend on the specific definition of $L_{TIR}$ and on the models 
used to compute it,  the $a_{MIR}$ coefficients for specific dust components or MIR broadband luminosities, which are easier to measure 
than the total IR, are independent of these choices. 

As stated above the smaller B04 corrections seem to be more appropriate than the usual $r$-band corrections given the agreement with data 
that do not require aperture corrections. However they are not trivially calculated (see B04 for details of the modeling).
More importantly \ha\ is often not easily measurable at all. It is therefore useful to provide SFR recipes based on a single IR 
quantity, or on a combination of IR and UV measurements (see next section) as UV is more easily obtained at high redshifts. 
Table \ref{table:coeffs} lists the mean  $L_{{\rm H}\alpha}^{corr}/L_{MIR}$ ratios of the SF population for the various MIR components.
Keeping in mind the non linearities and scatter in the true relations,
SFRs can be estimated from these approximate \ha\ luminosities using K09's calibration
(derived from the latest Starburst99 models and assuming a Kroupa IMF and solar metallicity):
\begin{equation}
SFR_{{\rm H}\alpha}~({\rm M_\odot~yr^{-1}})=5.5\times 10^{-42} ~L_{{\rm H}\alpha}^{corr}~({\rm ergs~s^{-1}}).
\label{eq:sfrha}
\end{equation}
As an example, the SFR derived from the MIPS 24\mi\ luminosity would be 
SFR$({\rm M_\odot~yr^{-1}})= 6.5\times 10^{-10} \nu L_{\nu}(24\mu m)/L_\odot$ consistent with Rieke et al. (2009) for
galaxies in the range of TIR luminosities of the present sample.

\begin{figure}
\plotone{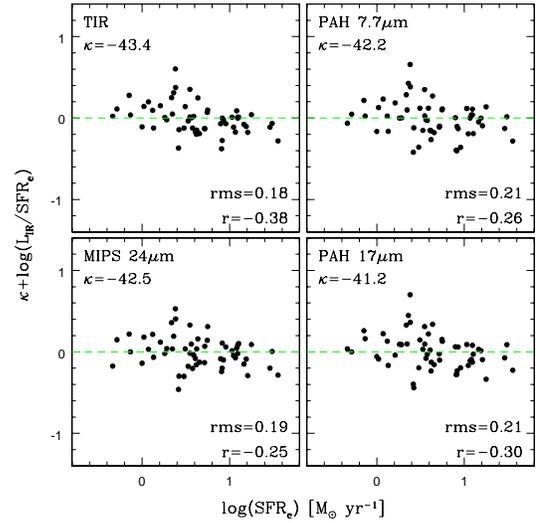}
\caption{$L_{IR}/SFR_e$ ratios as a function of \sfre\ (B04, see text for detail) where $L_{IR}$ equals the TIR, 
24\mi\ continuum, 7.7\mi\ and 17\mi\ PAH luminosities. $\kappa$ is defined as the median of log($SFR_e/L_{MIR}$). 
Only SF galaxies for which \sfre\ is computed from the Balmer lines are shown.  
\label{fig:sfre-sample}
}
\end{figure}

As Eq. \ref{eq:sfrha} was shown by B04 to underestimate the SFR of massive galaxies and thus may not be appropriate 
for this sample or at high $z$, 
we also add to Table \ref{table:coeffs} $SFR_e/L_{MIR}$ calibrations where \sfre\ is the SFR derived by these 
authors as follows: they computed SFR likelihood distributions of SF galaxies in the SDSS spectroscopic sample 
by fitting all strong emission lines simultaneously using the \cite{CharlotLonghetti01} 
models and assuming a Kroupa IMF. Dust was accounted for using the \cite{CharlotFall00} 
multicomponent model which provides a consistent treatment of the attenuation of both continuum and emission 
line photons. \sfre\ refers to the medians of these SFR distributions.
In this model, the \ha\ attenuation increases with mass while the ratio of $L_{{\rm H}\alpha}^{corr}$ 
to SFR decreases with mass so that the same observed \ha\ luminosity signals a noticeably higher SFR 
in higher mass galaxies than predicted from Kennicutt's relation. We refer to B04 for full details.
\sfre\ is found to be in good agreement with Eq. \ref{eq:sfrha} for average local galaxies but 
diverges from it for higher mass, higher metallicity galaxies such as found in the present sample where
\sfre\ is on average twice larger {\it within the SDSS fiber} than derived from the Kennicutt relation.
However the aperture corrections in this study being $\sim 1.6$ smaller than derived from the $r$-band 
magnitudes for SF galaxies, the total \sfre\ are only $\sim 1.3$ times larger than derived conventionally
using the Balmer decrements, $r$-band aperture corrrections and Eq. \ref{eq:sfrha}. 
For composite galaxies and AGNs, \sfre\ is not estimated from the emission lines which are contaminated by AGN 
emission, but in a statistical way based on the correlation between \myd4n\ and the specific SFR. We exclude
those for clarity.

Figure \ref{fig:sfre-sample} shows the relations between \sfre\ and the $L_{IR}/SFR_e$ ratios
for the TIR,  24\mi\ continuum, 7.7\mi\ and 17\mi\ PAH luminosities. 
As in previous figures the correlation parameters are quoted at the bottom right of each panel. 
These correlations are more scattered and less linear (higher rms and Pearson coefficient) than with $L_{H\alpha}^{corr}$.
The attenuations underlying \sfre\ being larger than those derived from the Balmer decrement for 
massive galaxies, the \sfre\ to \ltir\ ratio: 
$SFR_e=3.98\times 10^{-44}L_{TIR}$ is very close to that of Kennicutt et al. (1998) for opaque starburst galaxies
(taking into account the difference in IMFs). The $SFR_e/L_{MIR}$ calibrations are listed in Table \ref{table:coeffs}.

\subsection{MIR dust and UV}

Turning now to UV data where dust attenuation is an inevitable issue, we once again follow an energy balance
approach \citep{Meurer99, Gordon_etal2000, Kong04, Buat05, Cortese_etal2008, Zhu_etal2008, Kennicutt_etal2009}.
SFRs can be estimated from dust corrected FUV luminosities using the following calibration by K09 assuming a 
Kroupa IMF, solar metallicity, and adjusted to the GALEX FUV filter ($\lambda_{eff} = 1538$\AA).
\begin{equation}
SFR_{FUV}~({\rm M_\odot~yr^{-1}})=4.5\times 10^{-44} ~ L_{FUV}^{corr}~({\rm ergs~s^{-1}}).
\label{fig:sfruv}
\end{equation}
where $L_{FUV}^{corr}= \nu L_{\nu}^{corr} (1538{\rm \AA})$ is the dust-corrected GALEX FUV luminosity.

\begin{figure}
\plotone{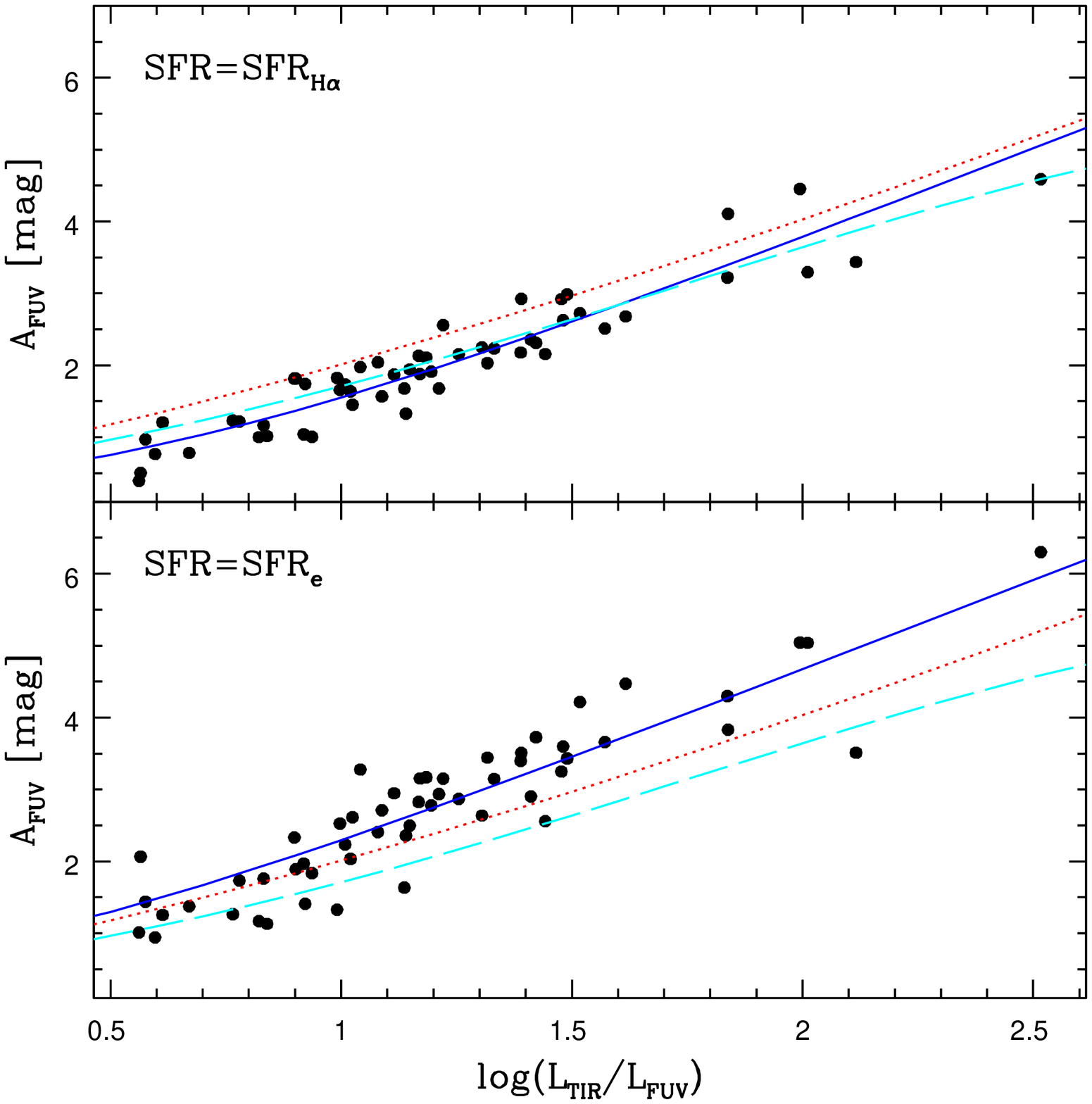}
\caption{The FUV attenuations of the SF population in the GALEX FUV band derived from Eq. \ref{eq:Afuv-sfr} as a function 
of $L_{TIR}/ \nu L_{\nu}^{obs}(1530\AA)$ (the IRX) assuming SFR $=SFR_{{\rm H}\alpha}$ (Eq. \ref{eq:sfrha}) and SFR=\sfre\ 
(top and bottom panels respectively). 
The dotted line is a theorical relation by \cite{Buat05}; the dashed lines shows a model derived 
by \cite{Cortese_etal2008} for galaxies with $FUV-g=2.9$ corresponding to the mean color of our sample; 
the solid lines are best fits to Eq. \ref{eq:Afuv-irx}.
\label{fig:Afuv-irx}
}
\end{figure}
\begin{figure}
\plotone{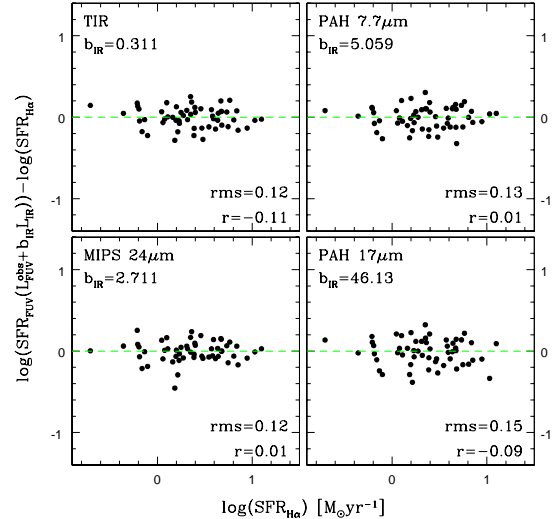}
\caption{The ratios of FUV to \ha\ SFRs against the \ha\ SFR: \ha\ is corrected using the Balmer decrement and the B04 aperture 
corrections while the FUV is dust corrected using Eqs. \ref{eq:Afuv-sfr} and \ref{eq:Afuv-irx} for the TIR, 24\mi\ continuum, 7.7\mi\ 
and 17\mi\ PAH luminosities. 
\label{fig:fuv-sample-ken}
}
\end{figure}

Assuming equality with a known SFR estimate (e.g. $SFR_{{\rm H}\alpha}$ or \sfre), we derive FUV attenuations as follows:
\begin{equation}
A_{FUV}	= 2.5~{\rm log}\left[ {SFR  \over SFR_{FUV}(L_{FUV}^{obs})} \right].
\label{eq:Afuv-sfr}
\end{equation}
where $L_{FUV}^{obs}=\nu L_{\nu}^{obs}(1538{\rm \AA})$ is the observed FUV luminosity in ${\rm ergs~s^{-1}}$.
Figure \ref{fig:Afuv-irx} shows the FUV attenuations of the SF subsample derived from Eq. \ref{eq:Afuv-sfr}
as a function of $L_{TIR}/ L_{FUV}^{obs}$ (known as the infrared excess or IRX) assuming assuming SFR=\sfrha\ (Eq. \ref{eq:sfrha}, top panel) 
and \sfre\ (bottom panel).
The median FUV attenuations are 1.9 and 2.8 magnitudes respectively, corresponding to $\sim$ 83 and 92\% of the FUV light being 
absorbed by dust (note that assuming conventional $r$-band aperture corrections for \ha\ yields exactly intermediate values). 
The dotted line is a theoretical relation by \cite{Buat05}; the dashed lines shows a model derived 
by \cite{Cortese_etal2008} for galaxies with $FUV-g=2.9$ corresponding to the mean color of our sample 
(these authors modeled the dependence of the IRX/$A_{FUV}$ relation with the age of the underlying stellar populations, 
or specific SFR, or color). The solid lines are best fits of the form: 
\begin{equation}
A_{FUV}	= 2.5~{\rm log}\left[ 1+ b_{IR} {L_{IR} \over L_{FUV}^{obs}} \right] 
\label{eq:Afuv-irx}
\end{equation}
equivalent to $L_{FUV}^{corr}= L_{FUV}^{obs}+ b_{IR} L_{IR}$, i.e.
$SFR= 4.5\times10^{-44} [L_{FUV}^{obs}+ b_{IR} L_{IR}]$, following K09's method.
Our best fit parameters are $b_{TIR}=0.317$ and 0.729 in the top and bottom panels respectively. However
all three models are poor in the bottom panel. FUV attenuations assuming \sfre\ are best modeled by a linear
function of log(IRX) or FUV-optical colors \citep{Treyer_etal2007}. 
Assuming \sfrha\ (top panel) the FUV attenuations are well fit both by \cite{Cortese_etal2008} and by Eq. \ref{eq:Afuv-irx}.
In this case a linear combination of $L_{FUV}^{obs}$ and \ltir\ or $L_{MIR}$ recovers \sfrha\ with low scatter as
shown in Figure \ref{fig:fuv-sample-ken} using the TIR, 24\mi\ continuum, 7.7\mi\ and 17\mi\ luminosities. 
As with \ha\ in the previous section, similarly good corrections can be achieved using other MIR 
components. The $b_{TIR}$ and $b_{MIR}$ coefficients are listed in Table \ref{table:coeffs}. 

\subsection{Neon emission lines}

\begin{figure} 
\plotone{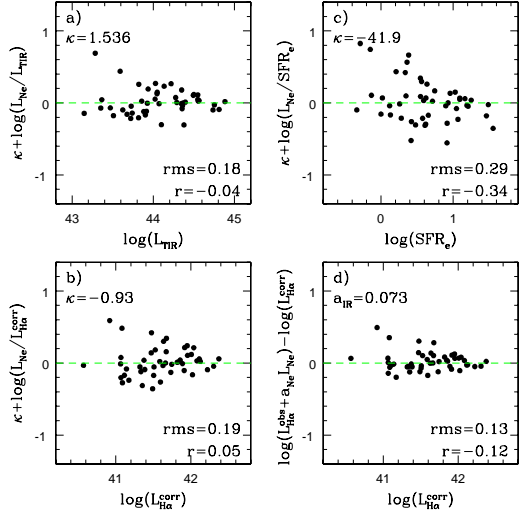}
\caption{The ratio of Ne luminosity (defined in Eq. \ref{eq:cloudy}) to \ltir\ (a), \lhacorr\ (b) and \sfre\ (c) 
as a function \ltir, \lhacorr\ and \sfre\ respectively. Only SF galaxies with measured metallicity are shown. 
The lower right panel (d) shows the ratio of the linear combination of \lhaobs\ and $L_{Ne}$ that best fits \lhacorr\ 
(see text for details) to \lhacorr\  against \lhacorr. 
\label{fig:neon}
}
\end{figure}

As put forward by \cite{HoKeto2007}, \neii\ is an excellent tracer of ionizing stars, being an abundant and dominant species 
in \hii\ regions, quite insensitive to density, as well as to dust given its long wavelength. \neiii\ has similar properties
but can be the dominant species in e.g. low-mass, low-metallicity galaxies \citep{OHalloran_etal2006, Wu_etal2006}. 
Thus Ne emission is expected to be directly comparable to the dust corrected \ha\ emission. 
Using the CLOUDY code \citep{CLOUDY1998}, we find that the ionizing flux from stars hotter than 10K is best 
represented by the following weighted linear combination of  \neii\ and \neiii\
also including a metallicity dependence:
\begin{equation}
{\rm H}\alpha = (8.8 {\rm [Ne~II] } + 3.5 {\rm [Ne~III] } ) \times (1/Z)^{0.8},
\label{eq:cloudy}
\end{equation}
where $Z$ is the metallicity in solar units. We use the right hand side of this equation to define the neon flux and luminosity, \lne.

For the sake of comparison with the study of \cite{HoKeto2007} who used \ltir\ as SFR estimate, as well as a straight sum of \neii\ and \neiii,
we note that our $L$(\neii+\neiii) to \ltir\ ratio is consistent with the IRS dataset used by these authors
\citep{OHalloran_etal2006, Wu_etal2006}. Our $L$(\neii) to \ltir\ ratio is larger but this may be 
explained by the large number of low metallicity galaxies in the samples used, in particular the \cite{Wu_etal2006}
dataset which specifically targets low-metallicity blue compact dwarf galaxies for which \neiii\ is the dominant
Ne species (cf. Figure \ref{fig:Ne-Z}).  

The left panels of Figure \ref{fig:neon} shows the \lne/\ltir\ and \lne/\lhacorr\ ratios as a function of \ltir\ (a) and \lhacorr\ (b)
respectively. 
Only SF galaxies with measured metallicity are represented (85\%). Surprisingly \lne\ behaves much like the MIR dust components.
It traces fairly linearly and tightly the total IR luminosity while  we can define $a_{Ne}=0.073\pm0.030$ using Eq. \ref{eq:Aha} such 
that  $L_{{\rm H}\alpha}^{obs}+ a_{Ne} L_{Ne}$ provides the tightest and most linear correlation with \lhacorr, as shown in
the lower right panel (d) of Figure \ref{fig:neon}. Likewise we can define $b_{Ne}=11.05\pm 5.13$ such that 
$4.5\times10^{-44} [L_{FUV}^{obs}+ b_{Ne} L_{Ne}]$ provides a good fit to \sfrha.
The upper right panel (c) shows the \lne\ to \sfre\ ratio against \sfre, which is significantly more
scattered than the previous relations as with the MIR dust components. This correlation implies the following calibration:
\begin{equation}
SFR({\rm M_\odot~yr^{-1}})=1.26\times 10^{-42} ~ L({\rm Ne})~({\rm ergs~s^{-1}}).
\end{equation}
The $a_{Ne}$ and $b_{Ne}$ coefficient as well as the median \ltir/\lne, \lhacorr/\lne \ and \sfre/\lne\ ratios are reported in Table \ref{table:coeffs}.

\subsection{Molecular Hydrogen lines}

\begin{figure}
\plotone{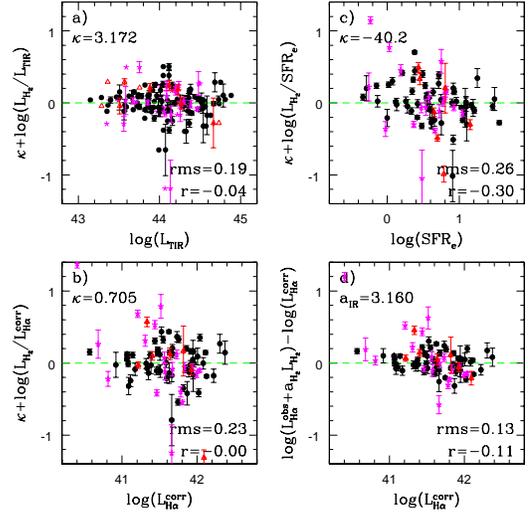}
\caption{The ratio of H$_2$ luminosity - defined as the sum of the $S(0)$ to $S(2)$ rotational lines of H$_2$ -  
to \ltir\ (a), \lhacorr\ (b) and \sfre\ (c) as a function \ltir, \lhacorr\ and \sfre\ respectively.
The lower right panel (d) shows the ratio of the linear combination of \lhaobs\ and $L_{{\rm H}_2}$ that best fits \lhacorr\ 
(see text for details) to \lhacorr\ against \lhacorr. 
\label{fig:H2}
}
\end{figure}

The rotational H$_2$ lines are fainter than the [NeII]12.9\mi, \neiii\ and \siii\ lines for most galaxies in our sample but 
molecular hydrogen represents a significant mass fraction of the ISM in normal galaxies.  The main excitation source of the rotational transitions 
is expected to be FUV radiation from massive stars in PDRs \citep[and references therein]{HollenbachTielens1997}, therefore these lines
also trace SF. The first study of warm molecular hydrogen ($T\sim 100 - 1000$ K) in the nuclei of normal, low luminosity galaxies 
was presented by \cite{Roussel_etal2007} (hereafter R07) using the SINGS sample. A major result of their work is the tight correlation 
between the sum of the $S(0)$ to $S(2)$ rotational lines (noted $F(S0-S2)$) and the PAH emission, 
with a $F(S0-S2)$/PAH ratio insensitive to the intensity of the radiation field. This correlation is interpreted 
as supporting the origin of H$_2$ excitation within PDRs (defined by \cite{HollenbachTielens1997} as including the neutral ISM illuminated by 
FUV photons), with fluorescence as the dominant excitation mechanism.

Our median logarithmic ratios of $L(S0-S2)$ to the TIR, 24\mi\ MIPS band and 7.7\mi\ PAH luminosities
for the SF population are $-3.17\pm 0.19$, $-2.18\pm 0.23$ and $-1.95\pm 0.19$ respectively.
The first two ratios are 1.6 and 1.8 times larger respectively than those of R07 for \hii\ nuclei 
(taking into account that R07 assumed a filter width of 3.1THz instead of $\nu F_{\nu}$ for the 24\mi\ band). 
On the other hand our $F(S0-S2)$/PAH ratio 
using the stellar component subtracted 8\mi\ IRAC flux instead of the PAHFIT extracted feature following R07, is only 1.2 times higher than that 
of R07. These differences are within uncertainties but the gradients may also reflect real differences between \hii\ nuclei and disks (warm
H$_2$ more abundant in disks), as well as support the physical link between warm H$_2$ and PAH emissions suggested by R07. 

Unlike R07 we do not find significantly higher $L(S0-S2)/L_{TIR}$ or  $L(S0-S2)/L_{24}$ ratios for AGNs (see also the top panel of Figure 
\ref{fig:lineratios}).
This may also be due to the much lower AGN contribution when disks are included. 
R07 interpret the higher AGN ratios as an excess of H$_2$ emission, attributed to additional mechanisms exciting H$_2$ molecules in AGNs. 
The $L(S0-S2)/L_{PAH}(7.7\mu m)$ ratio does show a significant excess for AGNs, however this excess correlates
with PAH EWs, suggesting that depleted PAHs in AGNs contribute in part to the effect. 
Our dispersion for the $L(S0-S2)/L_{PAH}(7.7\mu m)$ ratio is also comparable to the other two while R07 find it to be significantly tighter
in \hii\ nuclei, especially than the  $L(S0-S2)/L_{24}$ ratio. Our results suggest that local complexities are largely washed out on galactic scale 
and that warm molecular hydrogen traces dust in all its forms when considering integrated measurements. 

The left panels of Figure \ref{fig:H2} show the $L(S0-S2)$ to \ltir\ and \lhacorr\ ratios as a function of \ltir\ (a) and \lhacorr\ (b).
Like \lne, $L(S0-S2)$ traces reasonably linearly and tightly the total IR luminosity while the correlation with \lhacorr\ is improved with
a linear combination of $L_{{\rm H}\alpha}^{obs}$ and $L_{H_2}$ using Eq. \ref{eq:Aha} ($a_{H_2}=3.16\pm 2.74$) (c).
Likewise the upper right panel (d) shows the $L(S0-S2)$ to \sfre\ ratio against \sfre, which implies the following calibration:
\begin{equation}
SFR({\rm M_\odot~yr^{-1}})= 6.31\times 10^{-41} ~ L(S0-S2)~({\rm ergs~s^{-1}}).
\end{equation}
All the coefficients are reported in Table \ref{table:coeffs}.

\section{Summary and Conclusions}

We present a MIR spectroscopic survey of 100 `normal' galaxies at $z\sim 0.1$ with the goal
of investigating the use of mid-infrared PAH features, continuum and emission lines as probes of their 
star-formation and AGN activity. Available data include GALEX UV photometry, 
SDSS optical photometry and spectroscopy, and {\it Spitzer} near to far-infrared photometry. 
The optical spectroscopic data in particular allow us to classify these galaxies into star-forming, 
composite and AGNs, according to the standard optical ``BPT'' diagnotic diagram.
The MIR spectra were obtained with the low resolution modules of the {\it Spitzer} IRS and decomposed 
into unattenuated features and continuum using the PAHFIT code of \cite{Smith_etal2007}. 
A notable feature of this decomposition method is to extract a much larger PAH contribution
(and proportionally smaller continuum contribution) from the total flux compared to standard spline fitting methods 
which anchor the continuum in the wings of the features where non negligible PAH power remain. 
As a consequence, the PAH equivalenth widths are not only larger but extend over a considerably larger dynamic range
(e.g. the equivalent widths of the 6.2 and 7.7\mi\ PAH features in our sample extend to 15 and 32\mi\ 
respectively).

We study the variations of the various MIR spectral components as a function of the optically derived 
age (as measured by the 4000\AA\ break index), radiation field hardness (as measured by the
\myo3hb\ ratio) and metallicity (as measured by \n2ha\ ratio) of the galaxies. 
Systematic trends are found despite the lack of extreme objects in the sample, in particular
between PAH strength at low wavelength and gas phase metallicity, and between the ratio of high 
to low excitation lines (e.g. \oiv/\neii) and radiation field hardness. 
These trends confirm earlier results detected in sources with higher surface brightnesses such as ULIRGS, 
strong AGNs and \hii\ nuclei. Our results are consistent with the selective destruction in AGN radiation 
fields of the smallest PAH molecules efficient at producing the low wavelength PAH features (6.2 to 8.6\mi). 
They also suggest that radiation fields harder than those in the present sample would also destroy larger PAH 
molecules responsible for the longer wavelength features (11.3 to 17\mi). Aging galaxies also tend to
show weaker low wavelength PAH features, consistent with their main origin in star-forming regions.

We revisit the MIR diagnostic diagram of Genzel et al. (1998) relating PAH equivalent widths and 
\neii/\oiv\ emission line ratios. Based on the strongest trends we observed between these
measurements and optical emission line ratios and thanks to the extended range of equivalent widths 
provided by PAHFIT, we find this diagnostic to closely resemble the optical ``BPT'' diagram,
with a much improved resolving power for normal galaxies than previously found based on spline derived equivalent widths.
A mixed region of star-forming and composite galaxies remains, which may be revealing obscured AGNs 
in a large fraction of the optically defined `pure' star-forming galaxies.

We find tight and nearly linear correlations between the total infrared luminosity of star-forming galaxies and the luminosity 
of individual MIR components, including PAH features, continuum, neon emission lines and molecular hydrogen lines.
This implies that these individual MIR components are good gauges of the total dust emission on galactic scale
despite different spatial and physical origins on the scale of star-forming regions. Following the approach of \cite{Kennicutt_etal2009}
based on energy balance arguments, we show that like the total infrared luminosity, these individual components can 
be used to estimate dust attenuation in the UV and in the \ha\ lines. Given the non negligible attenuation in
these IR selected galaxies, the correlations between the MIR and dust corrected \ha\ luminosities can also
provide first order estimates of the SFR. We thus propose average scaling relations between the various
MIR components and \ha\ derived star-formation rates.


\clearpage

\begin{table*}
\caption{Absorption corrected continuum, PAH and line fluxes derived from the PAHFIT decompositions \label{table:fluxes} }
\begin{tabular}{lccccccccccc}
\tableline \tableline
\multicolumn{1}{c}{} & \multicolumn{3}{c}{continuum$^{(a)}$ [$10^{10}$ Jy Hz]} & \multicolumn{2}{c}{PAH [$10^{10}$ Jy Hz]} & \multicolumn{4}{c}{Lines [$10^{8}$ Jy Hz]} \\
\tableline
ID & 8\mi & 16\mi & 24\mi & 7.7\mi & 17\mi &    \neii & \neiii & \oiv  & H$_2$S(0)-S(2)$^{(b)}$ \\ 
\tableline
 1 & $      3.99 \pm     0.42 $ & $      6.07 \pm     0.37 $ & $     13.79 \pm     0.70 $ 	 & $      4.26 \pm     0.12 $ & $      0.48 \pm     0.02 $ 	& $     10.55 \pm      0.34 $ & $      5.11 \pm      0.30 $ & $      1.76 \pm      0.42  $ & $      5.48 \pm      0.61 $ \\ 
 2 & $      1.57 \pm     0.33 $ & $      2.21 \pm     0.26 $ & $      5.02 \pm     0.29 $ 	 & $      2.15 \pm     0.13 $ & $      0.42 \pm     0.03 $ 	& $      4.91 \pm      0.28 $ & $      2.40 \pm      0.33 $ & $      0.63 \pm      0.27  $ & $      3.12 \pm      0.58 $ \\ 
 3 & $      1.33 \pm     0.07 $ & $      1.53 \pm     0.06 $ & $      4.98 \pm     0.17 $ 	 & $      1.58 \pm     0.09 $ & $      0.18 \pm     0.03 $ 	& $      4.68 \pm      0.27 $ & $      1.56 \pm      0.25 $ & $      0.57 \pm      0.38  $ & $      6.69 \pm      0.65 $ \\ 
 4 & $      1.81 \pm     0.55 $ & $      3.60 \pm     0.96 $ & $      6.14 \pm     0.99 $ 	 & $      1.38 \pm     0.05 $ & $      0.41 \pm     0.03 $ 	& $      4.53 \pm      0.31 $ & $      2.02 \pm      0.23 $ & $      0.56 \pm      0.29  $ & $      3.42 \pm      0.49 $ \\ 
 5 & $      1.76 \pm     0.17 $ & $      2.54 \pm     0.12 $ & $      6.64 \pm     0.25 $ 	 & $      2.93 \pm     0.07 $ & $      0.45 \pm     0.06 $ 	& $      8.48 \pm      0.25 $ & $      1.96 \pm      0.25 $ & $      0.00 \pm      0.00  $ & $      4.69 \pm      1.73 $ \\ 
 6 & $      5.71 \pm     0.94 $ & $      8.04 \pm     1.89 $ & $     11.62 \pm     2.47 $ 	 & $      4.06 \pm     0.13 $ & $      0.94 \pm     0.05 $ 	& $     21.07 \pm      0.46 $ & $      1.81 \pm      0.24 $ & $      0.00 \pm      0.00  $ & $      7.41 \pm      0.51 $ \\ 
 7 & $      1.06 \pm     0.03 $ & $      2.63 \pm     0.10 $ & $      5.30 \pm     0.30 $ 	 & $      1.33 \pm     0.08 $ & $      0.28 \pm     0.02 $ 	& $      4.64 \pm      0.33 $ & $      1.11 \pm      0.21 $ & $      0.00 \pm      0.00  $ & $      2.39 \pm      0.49 $ \\ 
 8 & $      2.21 \pm     0.03 $ & $      5.47 \pm     0.07 $ & $      7.28 \pm     0.31 $ 	 & $      1.65 \pm     0.06 $ & $      0.49 \pm     0.02 $ 	& $      5.73 \pm      0.27 $ & $      1.63 \pm      0.34 $ & $      1.24 \pm      0.25  $ & $      6.64 \pm      0.42 $ \\ 
 9 & $      2.19 \pm     0.20 $ & $      2.26 \pm     0.06 $ & $      6.09 \pm     0.15 $ 	 & $      1.05 \pm     0.09 $ & $      0.17 \pm     0.02 $ 	& $      5.24 \pm      0.26 $ & $      0.16 \pm      0.20 $ & $      0.14 \pm      0.53  $ & $      1.99 \pm      0.43 $ \\ 
 10 & $      2.11 \pm     0.07 $ & $      1.44 \pm     0.09 $ & $      3.04 \pm     0.31 $ 	 & $      2.33 \pm     0.08 $ & $      0.38 \pm     0.02 $ 	& $      5.38 \pm      0.33 $ & $      2.52 \pm      0.30 $ & $      1.02 \pm      0.46  $ & $      3.85 \pm      0.41 $ \\ 
\tableline
\multicolumn{10}{l}{$(a)$ defined as $\nu F_{\nu}$;} \\
\multicolumn{10}{l}{$(b)$ the sum of H$_2$S(0) to S(2) lines.} \\
\end{tabular}
\end{table*}

\begin{table}
\caption{Aperture corrections in magnitudes \label{table:aper} }
\begin{tabular}{lccc}
\tableline \tableline
ID & IRAC 8\mi & IRSB 16\mi & MIPS 24\mi \\ 
\tableline 
1  &       0.34 &      0.09 &      0.06  \\ 
2  &       0.35 &     -0.01 &      0.08  \\ 
3  &       0.29 &      0.16 &      0.19  \\ 
4  &       0.86 &     -0.04 &     -0.08  \\ 
5  &       0.35 &      0.11 &      0.16  \\ 
6  &       0.55 &     -0.13 &     -0.10  \\ 
7  &       0.73 &      0.02 &      0.06  \\  
8  &       0.90 &     -0.18 &     -0.12  \\ 
9  &       0.54 &      0.23 &      0.25  \\ 
10  &       0.46 &      0.42 &      0.48  \\ 
\tableline
\end{tabular}
\end{table}

\begin{table}
\caption{PAH equivalent widths in $\mu m$ \label{table:eqw} }
\begin{tabular}{lccccc}
\tableline \tableline
ID & 6.2\mi & 7.7\mi & 8.6\mi & 11.3\mi  & 17\mi \\ 
\tableline 
 1 & $      2.99 \pm     0.38 $ & $      8.50 \pm     0.73 $ & $      1.80 \pm     0.22 $ & $      2.86 \pm     0.28 $ & $      1.19 \pm     0.12 $ \\ 
 2 & $      4.82 \pm     1.39 $ & $     11.30 \pm     2.06 $ & $      2.44 \pm     0.55 $ & $      3.07 \pm     0.47 $ & $      3.00 \pm     0.39 $ \\ 
 3 & $      2.91 \pm     0.27 $ & $      9.15 \pm     0.77 $ & $      1.90 \pm     0.18 $ & $      4.79 \pm     0.31 $ & $      1.71 \pm     0.29 $ \\ 
 4 & $      2.13 \pm     0.92 $ & $      6.37 \pm     1.60 $ & $      1.74 \pm     0.59 $ & $      2.08 \pm     0.53 $ & $      1.81 \pm     0.54 $ \\ 
 5 & $      2.59 \pm     0.57 $ & $     12.92 \pm     1.10 $ & $      3.00 \pm     0.31 $ & $      4.53 \pm     0.33 $ & $      2.70 \pm     0.43 $ \\ 
 6 & $      2.00 \pm     0.34 $ & $      5.67 \pm     0.68 $ & $      1.41 \pm     0.29 $ & $      1.99 \pm     0.36 $ & $      1.91 \pm     0.46 $ \\ 
 7 & $      2.55 \pm     0.44 $ & $     10.95 \pm     0.89 $ & $      2.13 \pm     0.16 $ & $      2.37 \pm     0.13 $ & $      1.64 \pm     0.17 $ \\ 
 8 & $      1.74 \pm     0.17 $ & $      6.48 \pm     0.30 $ & $      1.73 \pm     0.07 $ & $      2.09 \pm     0.06 $ & $      1.44 \pm     0.07 $ \\ 
 9 & $      0.93 \pm     0.27 $ & $      3.70 \pm     0.49 $ & $      0.67 \pm     0.10 $ & $      2.52 \pm     0.20 $ & $      1.09 \pm     0.13 $ \\ 
 10 & $      2.17 \pm     0.15 $ & $      8.48 \pm     0.45 $ & $      1.60 \pm     0.11 $ & $      4.27 \pm     0.20 $ & $      4.17 \pm     0.46 $ \\ 
\tableline
\end{tabular}
\end{table}


\begin{table}
\caption{Analytical boundaries in the plane of PAH equivalent widths [\mi] versus \neii/\oiv$^{(a)}$.  \label{table:diagnostics}}
\begin{tabular}{lrrrrrrrrrrr}
\tableline \tableline
\multicolumn{1}{c}{} & \multicolumn{3}{c}{6.2\mi} & \multicolumn{1}{c}{} &\multicolumn{3}{c}{7.7\mi} & \multicolumn{1}{c}{} & \multicolumn{3}{c}{8.6\mi} \\
\tableline
 & Eq. \ref{eq:mykewley} & Eq. \ref{eq:mykauffmann} & Eq. \ref{eq:myboundary}  &~~~& Eq. \ref{eq:mykewley} & Eq. \ref{eq:mykauffmann} & Eq. \ref{eq:myboundary}  &~~~& Eq. \ref{eq:mykewley} & Eq. \ref{eq:mykauffmann} & Eq. \ref{eq:myboundary}  \\
\tableline
$c_1$ &  2.36 & 1.96  & 1.00  &~~~& 2.21  & 1.36  & 1.30  &~~~& 1.84  & 1.10  & 1.20 \\
$c_2$ &  1.84 & 0.64  & 0.90  &~~~& 1.16  & -0.21 & 0.55  &~~~& 1.51  & 0.32  & 0.80 \\
$c_3$ & $-0.87$ & $-1.30$ & $-0.35$ &~~~& $-0.88$ & $-1.30$ & $-0.45$ &~~~& $-0.88$ & $-1.27$ & $-0.70$ \\
\tableline
\multicolumn{12}{l}{$(a)$ ${\rm log([NeII]/[OIV])} = [c_1/({\rm log(EW)}+c_2)]+c_3$.} \\
\multicolumn{12}{l}{~~~~ Eqs \ref{eq:mykewley}, \ref{eq:mykauffmann} and \ref{eq:myboundary} are defined in Section 3.3.} \\
\end{tabular}
\end{table}

\begin{table*}
\caption{Median ratios for the star-forming population \label{table:coeffs}}
\begin{tabular}{lcccccc}
\tableline \tableline
 & ${\rm log}\left[ L_{TIR}/L_{MIR}\right] $ & log$\left[ L_{H\alpha}^{corr}/ L_{MIR}\right]^{(a)} $ & log$\left[ L_{H\alpha}^{corr}/ L_{MIR}\right]^{(b)} $  & log$\left[SFR_e/ L_{MIR}\right]^{(c)}$ & $a_{IR}^{(d)}$ & $b_{IR}^{(e)}$ \\ 
\tableline 
TIR & $        0 \pm        0 $ & $     -2.266 \pm      0.159 $ & $     -2.471 \pm      0.144 $ & $    -43.461 \pm      0.183 $ & $      0.002 \pm      0.001 $ & $      0.318 \pm      0.124 $ \\
PAH 7.7\mi & $      1.204 \pm      0.087 $ & $     -1.091 \pm      0.181 $ & $     -1.242 \pm      0.154 $ & $    -42.248 \pm      0.213 $ & $      0.034 \pm      0.012 $ & $      5.154 \pm      2.299 $ \\
peak 7.7\mi & $      0.242 \pm      0.086 $ & $     -2.054 \pm      0.175 $ & $     -2.231 \pm      0.140 $ & $    -43.221 \pm      0.205 $ & $      0.004 \pm      0.001 $ & $      0.539 \pm      0.229 $ \\
PAH 11.3\mi & $      1.851 \pm      0.071 $ & $     -0.458 \pm      0.174 $ & $     -0.611 \pm      0.146 $ & $    -41.605 \pm      0.183 $ & $      0.143 \pm      0.053 $ & $     21.940 \pm      9.517 $ \\
PAH 17\mi & $      2.162 \pm      0.121 $ & $     -0.101 \pm      0.203 $ & $     -0.292 \pm      0.192 $ & $    -41.263 \pm      0.212 $ & $      0.320 \pm      0.160 $ & $     46.986 \pm     29.721 $ \\
IRAC 8\mi & $      0.594 \pm      0.075 $ & $     -1.719 \pm      0.177 $ & $     -1.896 \pm      0.142 $ & $    -42.871 \pm      0.195 $ & $      0.008 \pm      0.003 $ & $      1.214 \pm      0.514 $ \\
IRSB 16\mi & $      1.040 \pm      0.058 $ & $     -1.253 \pm      0.157 $ & $     -1.445 \pm      0.140 $ & $    -42.409 \pm      0.185 $ & $      0.022 \pm      0.007 $ & $      3.544 \pm      1.381 $ \\
MIPS 24\mi & $      0.959 \pm      0.096 $ & $     -1.311 \pm      0.142 $ & $     -1.505 \pm      0.151 $ & $    -42.508 \pm      0.191 $ & $      0.019 \pm      0.007 $ & $      2.763 \pm      1.253 $ \\
Ne$^{(f)}$ & $      1.537 \pm      0.183 $ & $     -0.738 \pm      0.216 $ & $     -0.930 \pm      0.199 $ & $    -41.901 \pm      0.298 $ & $      0.073 \pm      0.031 $ & $     11.258 \pm      5.228 $ \\
H$_2$S(0)-S(2)$^{(g)}$ & $      3.172 \pm      0.191 $ & $      0.888 \pm      0.255 $ & $      0.706 \pm      0.239 $ & $    -40.272 \pm      0.267 $ & $      3.161 \pm      2.641 $ & $    411.216 \pm    486.911 $ \\ 
\tableline
\multicolumn{7}{l}{$(a)$ using $r$-band aperture corrections} \\
\multicolumn{7}{l}{$(b)$ using B04 aperture corrections (see text for detail). Note that the relations in this case are markedly non-linear (right panel of Figure \ref{fig:lha-ltir}).} \\
\multicolumn{7}{l}{$(c)$ $SFR_e$ [M$_\odot$/yr] from B04 (see text for detail); $L_{MIR}$ [erg/s]}\\
\multicolumn{7}{l}{$(d)$ Eq. \ref{eq:Aha}}\\
\multicolumn{7}{l}{$(e)$ Eq. \ref{eq:Afuv-irx}}\\
\multicolumn{7}{l}{$(f)$ Eq. \ref{eq:cloudy} } \\
\multicolumn{7}{l}{$(g)$ the sum of H$_2$S(0) to S(2) lines} \\
\end{tabular}
\end{table*}

\end{document}